\documentclass[11pt]{article}

\usepackage[margin=.75in]{geometry}

\usepackage{amsmath,amssymb}
\usepackage{cancel}
\usepackage{graphicx}
\usepackage{caption}
\usepackage{comment}
\usepackage{subcaption}
\usepackage{color}
\definecolor{indigo}{RGB}{0,0,120}
\usepackage[colorlinks=true, linkcolor=indigo, citecolor=blue, urlcolor=indigo]{hyperref}

\newcommand{\Blue}{\color{blue}}

\def\tr{\,{\rm tr}\,}
\def\Tr{\,{\rm Tr}\,}
\def\sgn{\;{\rm sgn}\;}

\def\imply{\Rightarrow}

\def\fl{\noindent}

\newcommand{\tl}[1]{\tilde{#1}}
\newcommand{\dd}[2]{\frac {\partial #1}{\partial #2}}

\newcommand{\pdr}{\partial}

\newcommand{\beq}{\begin{equation}}
\newcommand{\eeq}{\end{equation}}
\newcommand{\beqs}{\begin{eqnarray}}
\newcommand{\eeqs}{\end{eqnarray}}
\newcommand{\half}{\frac{1}{2}}
\newcommand{\ov}[1]{\frac{1}{#1}}
\newcommand{\sign}[1]{{\rm sgn}({#1})}

\def\al{\alpha} 		
\def\del{\delta}
\def\D{\Delta}	
\def\g{\gamma} 
\def\eps{\epsilon} 
\def\la{\lambda}
		
\def\sig{\sigma}
\def\tht{\theta}
\def\om{\omega}		
\def\Om{\Omega}

\newcount\colveccount  
\newcommand*\colvec[1]{\global\colveccount#1  \begin{pmatrix} \colvecnext} \def\colvecnext#1{#1 \global\advance\colveccount-1
        \ifnum\colveccount>0 \\ \expandafter\colvecnext
        \else \end{pmatrix} \fi}

\usepackage{calligra} 
\DeclareMathAlphabet{\mathcalligra}{T1}{calligra}{m}{n}
\DeclareFontShape{T1}{calligra}{m}{n}{<->s*[2.2]callig15}{}
\newcommand{\scripty}[1]{\ensuremath{\mathcalligra{#1}}}

\begin{document}

\title{
\hfill {\tt \small \href{https://arxiv.org/abs/1906.03141}{arXiv:1906.03141[nlin.SI]}} \\
Invariant tori, action-angle variables and phase space structure of the Rajeev-Ranken model}
\author{{\sc Govind S. Krishnaswami and T. R. Vishnu}
\\ \small
Physics Department, Chennai Mathematical Institute,  SIPCOT IT Park, Siruseri 603103, India
\\ \small
Email: {\tt govind@cmi.ac.in, vishnu@cmi.ac.in}}
\date{August 10, 2019\\
Published in \href{https://doi.org/10.1063/1.5114668}{J. Math. Phys. {\bf 60}, 082902 (2019)}}

\maketitle

\abstract{ \small We study the classical Rajeev-Ranken model, a Hamiltonian system with three degrees of freedom describing nonlinear continuous waves in a 1+1-dimensional nilpotent scalar field theory pseudodual to the SU(2) principal chiral model. While it loosely resembles the Neumann and Kirchhoff models, its equations may be viewed as the Euler equations for a centrally extended Euclidean algebra. The model has a Lax pair and $r$-matrix leading to four generically independent conserved quantities in involution, two of which are Casimirs. Their level sets define four-dimensional symplectic leaves on which the system is Liouville integrable. On each of these leaves, the common level sets of the remaining conserved quantities are shown, in general, to be 2-tori. The non-generic level sets can only be horn tori, circles and points. They correspond to measure zero subsets where the conserved quantities develop relations and solutions degenerate from elliptic to hyperbolic, circular or constant functions. A geometric construction allows us to realize each common level set as a bundle with base determined by the roots of a cubic polynomial. A dynamics is defined on the union of each type of level set, with the corresponding phase manifolds expressed as bundles over spaces of conserved quantities. Interestingly, topological transitions in energy hypersurfaces are found to occur at energies corresponding to horn tori, which support purely homoclinic orbits. The dynamics on each horn torus is non-Hamiltonian, but expressed as a gradient flow. Finally, we discover a family of action-angle variables for the system that apply away from horn tori.}

\vspace{.25 cm}

\footnotesize

{\fl \bf Keywords:} Classical integrability, nonlinear waves, Euclidean algebra, nilpotent Lie algebra, Casimirs, symplectic leaves, common level sets, invariant tori, elliptic curve, gradient flow, action-angle variables.


\normalsize

{\scriptsize \tableofcontents}

\normalsize

\section{Introduction}

The Rajeev-Ranken model \cite{R-R, G-V} is a Hamiltonian system with three degrees of freedom. It arises as a reduction of a $1+1$-dimensional scalar field theory \cite{Z-M, Nappi} dual to the SU(2) principal chiral model (PCM) \cite{P-W}. Unlike the PCM, which is asymptotically free, the dual scalar field is strongly coupled in the ultraviolet and could serve as a toy-model to study non-perturbative features of theories with a Landau pole. This scalar field theory also arises as a large-level, weak-coupling limit of the Wess-Zumino-Witten model \cite{R-R}. In \cite{R-R}, the authors initiated the study of a class of highly nonlinear continuous waves in the scalar field theory, which could play a role similar to solitary waves in other field theories. The Rajeev-Ranken model is the reduction of the scalar field theory to the space of nonlinear screw-type waves of the form $\phi(x, t) = e^{Kx} R(t) e^{-Kx} + m K x$ satisfying the field equations $\ddot{\phi} = \phi''  + \la [\dot{\phi}, \phi']$ for the $\mathfrak{su}$(2) Lie algebra-valued scalar field $\phi$. Here, $K$ is a constant $\mathfrak{su}$(2) matrix, $m$ a dimensionless parameter  and $\la$ a dimensionless coupling constant. 

In \cite{R-R}, the evolution equations for $R(t)$ in the Rajeev-Ranken model were formulated as the ODEs 
	\beq
	\dot{\vec S}(t) = \la ( \vec S \times \vec L ) \quad \text{and} \quad 
	\dot{\vec L}(t) = \vec K \times \vec 
	S
	\label{e:EOM-RR}
	\eeq
for a pair of three-dimensional vectors $\vec L$ and $\vec S$ related to $R$ and $\dot R$ as in Eq. (\ref{e: L-and-S}). These equations were shown to admit a Hamiltonian formulation based on a quadratic Hamiltonian and a step-2 nilpotent Lie algebra mimicking those of the scalar field theory, with solutions expressible in terms of elliptic functions.  As we describe in Appendix \ref{a:Kirchhoff-equations}, the equations (\ref{e:EOM-RR}) may also be viewed as the Euler equations for a centrally extended Euclidean $\mathfrak{e}(3)$ algebra and the quadratic Hamiltonian $2 H = L^2 + (S - K/\la)^2$.

In \cite{G-V}, we studied the classical integrability of the Rajeev-Ranken model. We showed that it admits a family of degenerate but compatible Poisson structures, identified their Casimirs and found Lax pairs and $r$-matrices for the model. The model was argued to be Liouville integrable by displaying a complete set of four generically independent conserved quantities in involution.

The Rajeev-Ranken model is related to two other interesting dynamical systems: (a) In \cite{G-V}, a formal relation between its equations and those of the Neumann model \cite{B-T, B-B-T} was obtained. Though not an equivalence (as the corresponding dynamical variables live in different spaces), it was exploited to find a new Hamiltonian formulation for the Neumann model. (b) Interestingly,  Eqs. (\ref{e:EOM-RR}) also bear some resemblance to Kirchhoff's equations for a rigid body moving in an ideal potential flow \cite{MT}. Roughly, $\vec L$ and $\vec P = \vec S - \vec K/\la$ play the roles of total angular momentum and linear momentum of the body-fluid system in a body-fixed frame \cite{D-K-N}. However, while the Poisson brackets of the Kirchhoff system are given by the Euclidean $L$-$P$ Lie algebra, the Rajeev-Ranken model involves its central extension (see Appendix \ref{a:Kirchhoff-equations}).

The Rajeev-Ranken model may also be regarded as describing a special class of flat connections. Indeed, the currents  $r_0 = g^{-1}\dot g$ and $r_1 = g^{-1} g'$ of the PCM (for the SU(2) group-valued principal chiral field $g(x,t)$) are components of a flat $\mathfrak{su}$(2) connection in 1+1-dimensions, satisfying the additional condition $\dot{r_0}  = r_1'$. Solutions of the dual scalar field theory thus furnish a special class of flat connections $r_{\mu} = \la \epsilon_{\mu \nu} \pdr^{\nu} \phi$. There are other interesting integrable systems having to do with flat connections. For instance, in \cite{A-M, Audin, F-R} the authors study integrable systems describing Hamiltonian dynamics on the space of flat connections on a Riemann surface. Evidently, while solutions to the Rajeev-Ranken model are very special classes of flat connections, the latter models deal with evolution on the space of all flat connections.

In this article, we continue our investigations into the structure of the phase space and classical  integrability of the Rajeev-Ranken model. In \cite{G-V}, the question of finding all common level sets of conserved quantities and obtaining action-angle variables was posed. Here we find all common level sets and show that the phase space is foliated by four types of invariant tori [2-tori and their limits: horn tori (tori with equal major and minor radii - see Fig.~\ref{f:theta-phi-dynamics-3D-horn-torus}), circles and points]. Moreover, we show that the union of common level sets of a given type may be treated as the phase space of a self-contained dynamical system. We construct action-angle variables for the dynamics on the union of 2-tori, which occupy all but a measure zero subset of the phase space. They degenerate to action-angle variables on the union of circular level sets. Interestingly, we find that the dynamics on the space of horn tori is not Hamiltonian, but expressible as a gradient flow. We now summarize the organization of the paper along with the principal results of each section.

In section \ref{s:RR-model-introduction}, we introduce the Rajeev-Ranken model as a reduction of the 1+1-dimensional scalar field theory  and discuss its Hamiltonian formulation and classical integrability. 

In section \ref{s:redution-of-dynamics}, we use the conserved quantities $\mathfrak{c}, m, s$ and $h$ of the model to reduce the dynamics to their common level sets. To begin with, in section \ref{s:M4-cm-vectorfields-and-coordinates}, assigning numerical values to the Casimirs $\mathfrak{c}$ and $m$ of the nilpotent Poisson algebra of section \ref{s:RR-model-introduction}, enables us to reduce the six-dimensional degenerate Poisson manifold of the $S$-$L$ variables ($M^6_{S \text{-} L}$) to its non-degenerate four-dimensional symplectic leaves $M^4_{\mathfrak{c} m}$. We also find Darboux coordinates on $M^4_{\mathfrak{c} m}$ and use them to obtain a Lagrangian. Next, assigning numerical values to energy we find the generically three-dimensional energy level sets $M^E_{\mathfrak{c} m}$ and use Morse theory to discuss the changes in their topology as the energy is varied (see section \ref{s:Hill-region-Morse-theory}). Finally, in section \ref{s:Reduction-2D} we consider the common level sets $M^{s h}_{\mathfrak{c} m}$ of all four conserved quantities and argue that they are generically diffeomorphic to 2-tori. This is established by showing that they admit a pair of commuting tangent vector fields (the canonical vector fields $V_E$ and $V_h$ associated to the conserved energy and helicity $h$) that are linearly independent away from certain singular submanifolds. Section \ref{s:common-level-set-conserved-qtys} is devoted to a systematic identification of all common level sets of the conserved quantities $\mathfrak{c}, m, s$ and $h$. We find that the condition for a common level set to be nonempty is the positivity of a cubic polynomial $\chi(u)$, which also appears in the nonlinear evolution equation for $u = S_3/k$. Each common level set of conserved quantities may be viewed as a  bundle over a band of latitudes of the $S$-sphere $(\vec S \cdot \vec S = s^2 k^2)$, with fibres given by a pair of points that coalesce along the extremal latitudes (which must be zeros of $\chi$) (see Fig.~\ref{f:Common-level-set-L-space}). By analyzing the graph of the cubic $\chi$ (see Fig. \ref{f:Examples-of-common-level-set}) we show that the common level sets are compact and connected and can only be of four types: 2-tori (generic), horn tori, circles and single points (non-generic). The non-generic common level sets arise as limiting cases of 2-tori when the major and minor radii coincide, minor radius shrinks to zero or when both shrink to zero.

In section \ref{s:foliation-of-phase-space}, we study the dynamics on each type of common level set. The union of single point common level sets comprises the static subset: it is the union of a two and a three-dimensional submanifold ($\Sigma_2$ and $\Sigma_3$) of phase space. In section \ref{s:circular-level-sets}, we discuss the four-dimensional union $\cal C$ of all circular level sets. Circular level sets arise when $\chi$ has a double zero at a non polar latitude of the $S$-sphere. On $\cal C$, solutions reduce to trigonometric functions, the wedge product $dh \wedge ds^2 \wedge dm \wedge d\mathfrak{c}$ vanishes and the conserved quantities satisfy the relation $\D = 0$, where $\D$ is the discriminant of $\chi$. Geometrically, $\cal C$ may be realized as a circle bundle over a three-dimensional submanifold $\cal Q_{\cal C}$ of the space of conserved quantities. Finally, we find a set of canonical variables on $\cal C$ comprising the two Casimirs $\mathfrak{c}$ and $m$ and the action-angle pair $-kh$ and $\tht = \arctan(L_2/L_1)$.

In section \ref{s:horn-toroidal-level-sets}, we examine the four-dimensional union $\cal{\bar{H}}$ of horn toroidal level sets. It may be viewed as a horn torus bundle over a two-dimensional space of conserved quantities. Horn tori arise when the cubic $\chi(u)$ is positive between a simple zero and a double zero at a pole of the $S$-sphere. Solutions to the equations of motion degenerate to hyperbolic functions on $\cal{\bar{H}}$ and every trajectory is a homoclinic orbit which starts and ends at the center of a horn torus (see Fig.~\ref{f:theta-phi-dynamics-3D-horn-torus}). As a consequence, the dynamics on $\cal{\bar H}$ is not Hamiltonian, though we are able to express it as a gradient flow, thus providing an example of a lower-dimensional gradient flow inside a Hamiltonian system. Interestingly, though the conserved quantities are functionally related on horn tori, the wedge product $dh \wedge ds^2 \wedge dm \wedge d\mathfrak{c}$ is non-zero away from their  centers.

In section \ref{s:toroidal-level-sets}, we discuss the six-dimensional union $\cal T$ of 2-toroidal level sets, which may be realized as a torus bundle over the subset $\D \ne 0$ of the space of conserved quantities. We use two patches of  the local coordinates $\mathfrak{c}, m, h, s, \tht$ and $u$ to cover $\cal T$. The solutions of the equations of motion are expressed in terms of elliptic functions and the trajectories are generically quasi-periodic on the tori (see Fig~\ref{f:torus-plot-theta-phi}). By inverting the Weierstrass-$\wp$ function solution for $u$, we discover one angle variable. Next, by imposing canonical Poisson brackets, we arrive at a system of PDEs for the remaining action-angle variables, which remarkably reduce to ODEs. The latter are reduced to quadrature allowing us to arrive at a fairly explicit formula for a family of action-angle variables. In an appropriate limit, these action-angle variables are shown to degenerate to those on the circular submanifold $\cal C$. 

It is satisfying that a detailed  and explicit analysis of the dynamics and phase space structure of this model has been possible using fairly elementary methods. Our results should be helpful in understanding other aspects of the model's integrability (bi-Hamiltonian formulation on symplectic leaves, spectral curve etc.), the stability of its solutions, effects of perturbations and its quantization (for instance via our action-angle variables, through the representation theory of nilpotent Lie algebras or via path integrals using our Lagrangian obtained from Darboux coordinates, to supplement the Schr\"odinger picture results in \cite{R-R}). Quite apart from its physical origins and possible applications, we believe that the elegance of the Rajeev-Ranken model justifies a detailed study. It is hoped that the insights gained can then also be usefully applied to understanding the parent scalar field theory.

\section{Formulation of the classical Rajeev-Ranken model}
\label{s:RR-model-introduction}

The scalar field theory whose reduction leads to the Rajeev-Ranken model \cite{R-R} was introduced by Zakharov and Mikhaliov \cite{Z-M} and Nappi \cite{Nappi}. It is defined by the nonlinear field equations
	\beq
	\ddot{\phi} = \phi''  + \la [\dot{\phi}, \phi'] \quad \text{with dimensionless coupling} \quad \la > 0.
	\label{e:scalar-field-theory}
	\eeq 
Here, the traceless anti-hermitian matrix $\phi(x,t)$ is an $\mathfrak{su}(2)$ Lie algebra-valued scalar field. Using the screw-type (internal rotation and translation) continuous wave ansatz:
	\beq
	\phi(x,t) = e^{Kx} R(t) e^{-Kx} + m K x \quad \text{where} \quad K = \frac{i k \sigma_3}{2},
	\eeq
Eq. (\ref{e:scalar-field-theory}) reduces to a system of ODEs for a mechanical system with three degrees of freedom. Here $m$ and the wave number $k$ are real parameters. Unlike solitons, these nonlinear waves have constant energy density \cite{R-R}. Introducing the $\mathfrak{su}(2)$ matrices 
	\beq
	L = \left[ K, R \right] + mK
	\quad \text{and}
	\quad S = \dot{R} + \ov{\la} K,
	\label{e: L-and-S}
	\eeq
which play the roles of $\phi'$ and $\dot{\phi}$, the equations of motion (\ref{e:scalar-field-theory}) become six first order equations
	\beq
	\dot{L} = \left[K, S\right] 
	\quad \text{and} \quad 
	\dot{S} = \la \left[S, L\right]
	\label{e: EOM-LS}
	\eeq
for the components $L_a = i \tr(L \sig_a)$ and $S_a = i \tr(S \sig_a)$, where $\sig_a$ are the Pauli matrices. Here, $L_3 = -m k$ is non-dynamical. We will often use polar coordinates ($r, \tht, \rho, \phi$) for	
	\beq
	L_1 = kr \cos \tht, \quad L_2 = kr \sin \tht, \quad S_1 = k \rho \cos \phi \quad \text{and} \quad S_2 = k \rho \sin \phi
	\label{e:L-S-polar}
	\eeq
and work with the dimensionless variable $u= S_3/k$ in place of $S_3$. The equations of motion (\ref{e: EOM-LS}) of the Rajeev-Ranken model follow from the Hamiltonian
	\beq
	H = \half \Tr \left[ \left(S-\frac{1}{\la}K\right)^2 +  L^2 \right]
	=  \frac{S_a^2 + L_a^2}{2} + \frac{k}{\la} S_3 + \frac{k^2}{2\la^2} \quad ( \text{here} \; \Tr = -2 \tr) 
	\label{e: H-mechanical}
	\eeq
in conjunction with the step-2 nilpotent Poisson algebra
	\beq
	\left\{ L_a, L_b \right\}_{\nu} = 0, 
	\quad \left\lbrace S_a, S_b \right\rbrace_{\nu} = \la \epsilon_{abc} L_c
	\quad \text{and} \quad 
	\left\lbrace S_a, L_b \right\rbrace_{\nu} = -\epsilon_{abc} K_c,
	\label{e: PB-SL}
	\eeq
for $a,b = 1,2,3$. Interestingly, (\ref{e: EOM-LS}) also follow from the same Hamiltonian and the  distinct but compatible non-nilpotent Euclidean $\mathfrak{e}(3)$ Poisson algebra
	\beq
	\{ S_a, S_b \}_{\varepsilon} = 0, \quad \{ L_a, L_b \}_{\varepsilon} = - \la \eps_{abc} L_c \quad \text{and} \quad \{ L_a , S_b \}_{\varepsilon} = - \la \eps_{abc} S_c.
	\label{e:PB-SL-dual}
	\eeq
Both the nilpotent and Euclidean algebras are degenerate. Their centers are generated by the Casimirs $(\mathfrak{c}, m)$ and $(h,s)$ respectively, where	
	\beq
	\mathfrak{c} k^2 = \half L_a L_a + \frac{k}{\la} S_3, \quad
	m k^2 = - k L_3, \quad
	h k^2 = S_a L_a \quad \text{and} \quad
	s^2 k^2 = S_a S_a.
	\label{e:conserved-quantities}
	\eeq
In fact, there is a degenerate Poisson pencil's worth of brackets (see \S 4.2 of \cite{G-V})
	\beq
	\{ f, g \}_{\al} = (1 - \al) \{ f, g \}_{\nu} + \al \{ f, g \}_{\epsilon},
	\label{e:Poisson-pencil}
	\eeq 
all of which imply (\ref{e: EOM-LS}) with the same Hamiltonian (\ref{e: H-mechanical}). Henceforth, we work with the nilpotent Poisson structure so that $s$ and $h$ are  non-Casimir conserved quantities.	
The Hamiltonian (\ref{e: H-mechanical}) can be expressed as 
	\beq
	H = k^2 E =  k^2 \left(\frac{s^2}{2}+ \mathfrak{c} + \ov{2\la^2} \right).
	\label{e:Hamiltonian-s}
	\eeq
The equations of motion (\ref{e: EOM-LS}) were shown \cite{G-V} to be equivalent to the Lax equation $\dot A = [B, A]$ with spectral parameter $\zeta$ where
	\beq
	A(\zeta) = -K\zeta^2 + L\zeta + \frac{S}{\la} 
	\quad \text{and} \quad	
	B(\zeta) = \frac{S}{\zeta}.
	\label{e:Lax-pair}
	\eeq	
The conserved quantities $\mathfrak{c}, m, s$ and $h$ arise as coefficients of  powers of $\zeta$ in $\Tr A^2$. They are in involution due to the existence of an $r$-matrix:
	\beq
	\left\lbrace A(\zeta) \stackrel{\otimes}{,} A(\zeta') \right\rbrace_\nu = \left[ r(\zeta, \zeta') , A(\zeta)\otimes I + I \otimes A(\zeta')\right]  \quad \text{where} \quad
	r(\zeta, \zeta') = - \frac{P}{2\la (\zeta - \zeta')}. 
	\label{e:r-matrix-nilpotent}
	\eeq
Here, $P$ is the permutation matrix. In \cite{G-V}, it was shown that $\mathfrak{c}, m, s$ and $h$ are a complete and generically independent set of conserved quantities. Along with their involutive property, this was used to argue that the dynamics on each four-dimensional symplectic leaf (obtained by fixing the values of the two Casimirs $\mathfrak{c}$ and $m$) must be Liouville integrable.

\section{Using conserved quantities to reduce the dynamics}
\label{s:redution-of-dynamics}

In this section, we discuss the reduction of the six-dimensional $S$-$L$ phase space ($M^6_{S \text{-} L}$) by successively assigning numerical values to the conserved quantities $\mathfrak{c}, m, s$ and $h$. For each value of the Casimirs $\mathfrak{c}$ and $m$ we obtain a four-dimensional manifold $M^4_{\mathfrak{c} m}$ with non-degenerate Poisson structure, which is expressed in local coordinates along with the equations of motion. Next, we identify the (generically three-dimensional) constant energy submanifolds $M^E_{\mathfrak{c} m} \subset M^4_{\mathfrak{c} m}$, where $E$ is a function of $s$ and $\mathfrak{c}$ (\ref{e:Hamiltonian-s}). Moreover, we use Morse theory to study the changes in topology of $M_E^{\mathfrak{c} m}$ with  changing energy. Finally, the conservation of helicity $h$ allows us to reduce the dynamics to generically two-dimensional manifolds $M^{s h}_{\mathfrak{c} m} $, which are the common level sets of all four conserved quantities. By analysing the nature of the canonical vector fields $V_{E}$ and $V_h$, the latter are shown to be 2-tori in general. We also argue that there cannot be any additional independent integrals of motion. Though the common level sets of all four conserved quantities $M_{\mathfrak{c} m}^{s h}$ are generically 2-tori, there are other possibilities. We show that $M_{\mathfrak{c} m}^{s h}$ has the structure of a bundle over a portion of the sphere $\Tr S^2 = s^2 k^2$, determined by the zeros of a cubic polynomial $\chi(u)$. By analyzing the possible graphs of $\chi$ we show that $M_{\mathfrak{c} m}^{s h}$ is compact, connected and of four possible types: tori, horn tori, circles and points. 

\subsection{Using Casimirs $\mathfrak{c}$ and $m$ to reduce to 4D phase space $M^4_{\mathfrak{c} m}$}
\label{s:M4-cm-vectorfields-and-coordinates}
\subsubsection{Symplectic leaves $M^4_{\mathfrak{c} m}$ and energy and helicity vector fields}
\label{s:canonical-vector-fields}

The common level sets of the Casimirs $\mathfrak{c}$ and $m$ are the four-dimensional symplectic leaves $M^4_{\mathfrak{c} m} \cong \mathbb{R}^4$   of the phase space $M^6_{S \text{-} L}$. On $M^4_{\mathfrak{c} m}$, the Poisson tensor $\scripty{r}^{ab}$ corresponding to the nilpotent Poisson algebra (\ref{e: PB-SL}) (see also \S 4.1 of \cite{G-V}) is non-degenerate and may be inverted to obtain the symplectic form $\omega_{a b}$. In Cartesian coordinates $\xi^a = (L_1, L_2, S_1, S_2)$,
	\beq
	\scripty{r}^{a b} = i k \colvec{2}{0 & \sig_2}{\sig_2 & -\la m \sig_2}
	\quad \text{and} \quad
	\omega_{a b} = (\scripty{r}^{-1})_{a b} = -\frac{i}{k}\colvec{2}{m \la \sig_2 &  \sig_2}{\sig_2 & 0}.
	\eeq
This symplectic form $\omega = (1/2) \omega_{ab} d\xi^a \wedge d\xi^b$ is the exterior derivative of the canonical 1-form $\al = -(1/2) \omega_{ab} \xi^b d\xi^a$. Expressing helicity $h$ (\ref{e:conserved-quantities}) and $E$ (\ref{e:Hamiltonian-s}) as functions on $M^4_{\mathfrak{c} m}$ by eliminating 
	\beq
	S_3 (L_1, L_2) = \frac{\la k}{2} \left( \left(2\mathfrak{c} -m^2\right) -\ov{k^2}(L_1^2 + L_2^2) \right) 
	\quad \text{and} \quad 
	L_3 = -m k
	\label{e:S3-L12}
	\eeq
we obtain the helicity and Hamiltonian vector fields on $M^4_{\mathfrak{c} m}$ (see also Eq. (61) of \cite{G-V})
	\beqs
	k V_h &=&  L_2 \pdr_{L_1} - L_1 \pdr_{L_2} + S_2 \pdr_{S_1} - S_1 \pdr_{S_2} \quad \text{and} \cr
	k V_E &=&  S_2 \pdr_{L_1} - S_1 \pdr_{L_2} - \left[\la \frac{S_3 L_2}{k} + \la m S_2 \right] \pdr_{S_1} + \left[\la \frac{S_3 L_1}{k} + \la m S_1 \right] \pdr_{S_2}.
	\label{e:Hamiltonian-vector-fields}
	\eeqs
Since $E$ and $h$ commute, $\omega(V_E, V_h) = \{ E, h \} = 0$. It is notable that $V_h$ is non-zero except at the origin ($S_{1,2} = L_{1,2} = 0$), while $V_E$ vanishes at the origin and on the circle ($L_1^2 + L_2^2 = k^2 (2 \mathfrak{c} - m^2), S_{1,2} = 0$). The points where $V_E$ and $V_h$ vanish turn out be the intersection of $M^4_{\mathfrak{c} m}$ with the static submanifolds  
	\beq
	\Sigma_{2} = \{ \vec S, \vec L \:|\: S_{1,2} = L_{1,2} = 0 \} \quad \text{and} \quad 
	\Sigma_3 = \{ \vec S, \vec L \:|\: \vec S = 0 \}
	\label{e:static-submanifolds}
	\eeq
introduced in \S 5.5 of \cite{G-V}, where  $S$ and $L$ are time-independent. The points where $V_E$ vanish will be seen in \S \ref{s:Hill-region-Morse-theory} to be critical points of the energy function. 

\subsubsection{ Darboux coordinates on symplectic leaves $M^4_{\mathfrak{c} m}$}

Since $M^4_{\mathfrak{c} m} \cong \mathbb{R}^4$ it is natural to look for global canonical coordinates. In fact, the canonical coordinates $(R_a, k P_b)$ on the six-dimensional phase space $M^6_{R \text{-} P}$ (which were introduced in  \S 4.3 of \cite{G-V}) restrict to Darboux coordinates on $M^4_{\mathfrak{c} m}$:
	\beq
	k R_a = - \eps_{ab}L_b \quad \text{and} \quad
	k P_a = S_a + \frac{\la m}{2} L_a \quad \text{for} \quad a,b = 1,2
	\label{e:R-P-relation-to-L-S}
	\eeq
with $\{ R_a, k P_b \} = \del_{ab}$ and $\{ R_a, R_b \} = \{ P_a, P_b \} = 0$. The Hamiltonian is a quartic function in these coordinates:
	\beq
	\frac{H}{k^2} = \frac{ P_1^2 + P_2^2}{2} + \frac{\la m}{2}(R_1 P_2 -R_2 P_1) + \frac{\la^2}{8} (R_1^2 + R_2^2) \left(R_1^2 + R_2^2 + 3 m^2 - 4 \mathfrak{c} \right) + \frac{\la^2}{8} (2 \mathfrak{c} - m^2)^2 + \mathfrak{c} + \frac{1}{2 \la^2}.
	\eeq
The equations of motion resulting from these canonical Poisson brackets and Hamiltonian are cubically nonlinear ODEs. In fact, for $a = 1, 2$:
	\beq
	k^{-1} \dot{R_a} = P_a - \frac{\la m}{2} \eps_{a b} R_b \quad \text{and} \quad
	k^{-1} \dot{P_a} = -\frac{\la m}{2} \eps_{a b} P_b - \frac{\la^2}{4} \left(3 m^2  - 4 \mathfrak{c} + 2 R_b R_b \right)R_a.
	\eeq
A Lagrangian $L_{\mathfrak{c} m}(R, \dot R)$,  leading to these equations of motion can be obtained by extremizing $k P_a \dot{R_a} - H$ with respect to $P_1$ and $P_2$:
	\beqs
	L_{\mathfrak{c} m} &=& \half \left( \dot{R_1}^2 + \dot{R_2}^2 - \la m k (R_1 \dot{R_2} - R_2 \dot{R_1}) \right) - \frac{\la^2 k^2}{8} (R_1^2 + R_2^2) \left(R_1^2 + R_2^2 + 2 m^2 - 4 \mathfrak{c} \right) \cr
	&&- k^2 \left(\frac{\la^2}{8} (2 \mathfrak{c} - m^2)^2 + \mathfrak{c} + \frac{1}{2 \la^2} \right).
	\eeqs

\subsection{Reduction to tori using conservation of energy and helicity}
\label{s:Reduction-2D}

So far, we have chosen (arbitrary) real values for the Casimirs $\mathfrak{c}$ and $m$ to arrive at the reduced phase space $M^4_{\mathfrak{c} m}$. Now assigning numerical values to the Hamiltonian $H = E k^2$ we arrive at the generically three-dimensional constant energy submanifolds $M^E_{\mathfrak{c} m}$ which foliate $M^4_{\mathfrak{c} m}$. It follows from the formula for the Hamiltonian (\ref{e:Hamiltonian-s}) that each of the $S_a$ is bounded above in magnitude by $|k|s = \sqrt{2 k^2 (E - \mathfrak{c} - 1/2 \la^2)}$. Moreover, $M^E_{\mathfrak{c} m}$ is closed as it is the inverse image of a point. Thus, constant energy manifolds are compact. Interestingly, the topology of $M^E_{\mathfrak{c} m}$ can change with energy: this will be discussed in \S \ref{s:Hill-region-Morse-theory}. In addition to the Hamiltonian and Casimirs $\mathfrak{c}$ and $m$, the helicity $h k^2 = \Tr SL$ is a fourth (generically independent) conserved quantity (see \S 5.7 of \cite{G-V}). Thus each trajectory must lie on one of the level surfaces $M^{E h}_{\mathfrak{c} m}$ of $h$ that foliate $M^E_{\mathfrak{c} m}$. Note that since $s \geq 0$ is uniquely determined by $E$ (and vice versa), the level sets of the conserved quantities $M^{E h}_{\mathfrak{c} m}$ and $M^{s h}_{\mathfrak{c} m}$ are in 1-1 correspondence and we will use the two designations interchangeably. 

We will see in \S \ref{s:canonical-vector-fields-topology} that these common level sets  of conserved quantities $M^{E h}_{\mathfrak{c} m}$ are generically 2-tori, parameterized by the angles $\tht$ and $\phi$ which (as shown in \S 5.2 of \cite{G-V}) evolve according to
	\beq
	\dot{\theta} =  -k\left(\frac{h + m u}{2\mathfrak{c} - m^2 - 2u/\la}\right) \quad \text{and} \quad
	\dot{\phi} = k m \la + k \la u \left(\frac{h + mu}{s^2 - u^2}\right).
	\label{e:theta-phi-dynamics}
	\eeq
Here, $u = S_3/k$ is related to $\tht$ and $\phi$  via helicity $h k^2 = \Tr S L$ and (\ref{e:conserved-quantities})
	\beq
	\sqrt{\left(s(E,\mathfrak{c})^2-u^2 \right) \left(2\mathfrak{c} - m^2 - 2u/\la \right)} \: \cos(\theta - \phi) = h + mu.
	\label{e:relation-theta-phi-u}
	\eeq
In other words, the components $V_E^\tht = \dot{\theta}/k^2$ and $V_E^\phi = \dot{\phi}/k^2$ of the Hamiltonian vector field $V_E = V_E^\tht \pdr_\tht + V_E^\phi \pdr_\phi$ are functions of $u$ alone. Though the denominators in (\ref{e:theta-phi-dynamics}) could vanish, the quotients exist as limits, so that $V_E$ is non-singular on $M^{s h}_{\mathfrak{c} m}$. Interestingly, as pointed out in \cite{R-R}, $u$ evolves by itself as we deduce from (\ref{e: EOM-LS}):
	\beq
	\dot{u}^2 = \la^2  k^2 \rho^2 r^2 \sin^2(\theta - \phi) = \la^2  k^2 \left[ (s^2 - u^2)\left(2 \mathfrak{c} - m^2 - \frac{2u}{\la}\right) - (h + mu)^2\right] = 2 \la k^2 \chi(u).
	\label{e:EOM-u}
	\eeq
This cubic $\chi(u)$ will be seen to play a central role in classifying the invariant tori in \S \ref{s:common-level-set-conserved-qtys}. The substitution $u = a v + b$, reduces this ODE to Weierstrass normal form
	\beq
	\dot v^2 = 4 v^3 - g_2 v - g_3, \quad \text{where} \quad a = 2/k^2 \la \quad \text{and} \quad  b = \mathfrak{c} \la/ 3 
	\label{e:Weierstrass-normal-form}
	\eeq
with solution $v(t) = \wp(t + \alpha; g_2, g_3)$. Here, the Weierstrass invariants are (there is a minor error in $g_3$ in \cite{R-R})
	\beq
	g_2 = \frac{ k^4 \la^2}{3} (3 \la h m + \la^2 \mathfrak{c}^2 + 3 s^2), \quad
	g_3 = \frac{k^6 \la^4}{108}  (27 h^2 + 18 \la \mathfrak{c} m h + 4 \la^2 \mathfrak{c}^3 - 36 \mathfrak{c} s^2 + 27 m^2 s^2).
	\label{e:Weierstrass-invariants}
	\eeq
Thus we obtain
	\beq
	u(t) = \frac{2}{k^2 \la} \wp(t + \alpha) + \frac{\mathfrak{c} \la}{3}
	\label{e:u-wp-function}
	\eeq
which oscillates periodically in time between $u_{\rm min}$ and $u_{\rm max}$, which are neighbouring zeros of $\chi$ between which $\chi$ is positive. Choosing $\al$ fixes the initial condition, with its real part fixing the origin of time. In particular, if $\al =\omega_I$ (the imaginary half-period of $\wp$), then $u(0) = u_{\rm min}$. On the other hand, $u(0) = u_{\rm max}$ if $\al = \om_R + \om_I$, where $\om_R$ is the real half-period. The formula (\ref{e:u-wp-function}) will be used in \S \ref{s:toroidal-level-sets} to find a set of action-angle variables for the system.

\subsubsection{Reduction of canonical vector fields to $M_{\mathfrak{c} m}^{s h}$ and its topology}
\label{s:canonical-vector-fields-topology}

In this section, we use the coordinates $(s^2, h , \tht, \phi)$ to show that the canonical vector fields $V_E$ and $V_h$ are tangent to the level sets $M^{s h}_{\mathfrak{c} m}$, which are shown to be compact connected Lagrangian submanifolds of the symplectic leaves $M^4_{\mathfrak{c} m}$. Moreover, $V_E$ and $V_h$ are shown to be generically linearly independent and to commute, so that $M^{s h}_{\mathfrak{c} m}$ are generically 2-tori.

On $M^4_{\mathfrak{c} m}$, the coordinates $(s^2, h , \tht, \phi)$ (as opposed to $(L_1,L_2,S_1,S_2)$) are convenient since the common level sets $M^{s h}_{\mathfrak{c} m} \subset M^4_{\mathfrak{c} m}$ arise as intersections of the $s^2$ and $h$ coordinate hyperplanes. The remaining variables $\tht$ and $\phi$ furnish coordinates on $M^{s h}_{\mathfrak{c} m}$. The Poisson tensor on $M^4_{\mathfrak{c} m}$ in these coordinates has a block structure, as does the symplectic form:
	\beq
	\scripty{r}^{a b} = \ov k \colvec{2}{0 & \al}{- \al^t & \beta} \quad \text{and} \quad \omega_{a b} = k \colvec{2}{-\g & -\delta^t}{ \delta & 0},
	\label{e:r-omega-s-h-tht-phi-coordinate}
	\eeq
where $\al, \beta, \g$ and $\delta$ are the dimensionless $2 \times 2$  matrices:
	\beqs
	\al &=& \colvec{2}{ \frac{-2 \dot \tht}{k} & \frac{-2 \dot \phi}{k}}{1 & 1}, \quad \beta =-i\frac{s_{\tht \phi}}{r \rho} \sigma_2, \quad \g = \left(- \al^t\right)^{-1} \beta \alpha^{-1} = -\frac{\beta}{\det \alpha} \cr
	 \text{and} \quad \delta &=& \al^{-1} = \frac{1}{\det \al}\colvec{2}{1 & \frac{2 \dot \phi}{k} }{-1 & \frac{-2 \dot \tht}{k}} \quad \text{with} \quad 
	 \det \al = k^2\sqrt{\det \scripty{r}} = \frac{-2}{k} \left(\dot \tht - \dot \phi \right).
	\eeqs
Here $s_{\tht \phi } = \sin(\tht - \phi)$ and $\dot \tht$ and $\dot \phi$ are as in (\ref{e:theta-phi-dynamics}), subject to the relation (\ref{e:relation-theta-phi-u}). From (\ref{e:conserved-quantities}), it follows that $\rho$ and $r$ may be expressed in terms of $s^2, h, \tht$ and $\phi$, by solving the pair of equations
	\beq
	h = r \rho c_{\tht \phi} - \frac{\la m}{2} \left( 2\mathfrak{c} - (r^2 + m^2) \right) \quad \text{and} \quad
	s^2 = \rho^2 + \frac{\la^2}{4} \left(2 \mathfrak{c} - (r^2 + m^2)\right)^2.
	\eeq 
Here $c_{\tht \phi } = \cos(\tht - \phi)$. In these coordinates, $V_h$ and $V_E$ (\ref{e:Hamiltonian-vector-fields}) have no components along $\pdr_s$ or $\pdr_h$: 
	\beq
	k V_h = -(\pdr_\tht + \pdr_\phi) \quad \text{and} \quad
	k V_{E} = -\frac{\rho}{r} c_{\tht \phi} \pdr_{\tht}  + \left(\la m + \frac{\la^2 r}{ 2 \rho} c_{\tht \phi} \left(2 \mathfrak{c} - (r^2 + m^2)\right) \right) \pdr_{\phi}.
	\label{e:hamiltonian-vector-field-s-h-tht-phi}
	\eeq
Thus, $V_h$ and $V_E$ are tangent to $M^{s h}_{\mathfrak{c} m}$. Moreover, the restriction of $\omega$ to $M^{s h}_{\mathfrak{c} m}$ is seen to be identically zero as it is given by the $\tht$-$\phi$ block in (\ref{e:r-omega-s-h-tht-phi-coordinate}) so that $M^{s h}_{\mathfrak{c} m}$ is a Lagrangian submanifold. Trajectories on $M^{s h}_{\mathfrak{c} m}$ are the integral curves of $V_E$.

To identify the topology of the common level set $M^{s h}_{\mathfrak{c} m}$, it is useful to investigate the linear independence (over the space of functions) of the vector fields $V_E$ and $V_h$. On $M^4_{\mathfrak{c} m}$, $\omega$ is non-degenerate so that $V_E$ and $V_h$ are linearly independent iff $dE \wedge dh \neq 0$. We find that this wedge product vanishes on $M^4_{\mathfrak{c} m}$ precisely when $S_{1,2}$ and $L_{1,2}$ satisfy the relations
	\beq
	\Xi_1: \; (S \times L)_3 = 0, \quad
	\Xi_2: -\la L_1(S \times L)_2 = k S_1^2 
	\quad \text{and} \quad
	\Xi_3: \: \la L_2 (S \times L)_1 = k S_2^2.
	\label{e:trigonometric-submanifold-conditions}
	\eeq
Here $(S \times L)_3 = S_1 L_2 - S_2 L_1$ etc., and $S_3$ and $L_3$ are expressed using (\ref{e:S3-L12}). It was shown in \S 5.6 of \cite{G-V} that (\ref{e:trigonometric-submanifold-conditions}) are the necessary and sufficient conditions for the four-fold wedge product $dh \wedge ds^2 \wedge dm \wedge d\mathfrak{c}$ to vanish in  $M^6_{S\text{-}L}$. Moreover, it was shown that this happens precisely on the singular set $\bar {\cal C} \subset M^6_{S\text{-}L}$ which consists of the circular/trigonometric submanifold ${\cal C}$ and its boundaries ${\cal C}_{1,2}$  and $\Sigma_{2,3}$. Thus, $V_E$ and $V_h$ are linearly independent away from the set (of measure zero) given by the intersection of $\bar{\cal C}$ with $M^4_{\mathfrak{c} m}$. [For given $\mathfrak{c}$ and $m$, the intersection of $\cal C$ with $M^4_{\mathfrak{c} m}$ is in general a two-dimensional manifold defined by four conditions among the six variables $\vec S$ and $\vec L$: $\Xi_1$ and $\Xi_2$ (with $S_{1,2} \neq 0$) as well as the conditions in Eq. (\ref{e:S3-L12}).] Furthermore, since $E$ and $h$ Poisson commute, $[V_E, V_h] = - V_{\{E,h \}} = 0$. {\it So, as long as we stay away from these singular submanifolds, $V_E$ and $V_h$ are a pair of commuting linearly independent vector fields tangent to $M^{s h}_{\mathfrak{c} m}$ (see Lemma 1 in Chapter 10 of {\cite{Arnold}}).} Additionally, we showed at the beginning of \S \ref{s:Reduction-2D}  that the energy level sets $M^E_{\mathfrak{c} m} \subset M^4_{\mathfrak{c} m}$ are compact manifolds. Now, $M^{s h}_{\mathfrak{c} m}$ must also be compact as it is a closed subset of $M^E_{\mathfrak{c} m}$ (the inverse image of a point). Finally, we will show in \S\ref{s:Examples-of-CL-sets-of-conserved-quantities} that $M^{s h}_{\mathfrak{c} m}$ is connected. Thus, for generic values of the conserved quantities, $M_{\mathfrak{c} m}^{s h}$ is a compact, connected surface with a pair of linearly independent tangent vector fields. By Lemma 2 in Chapter 10 of {\cite{Arnold}}, it follows that the common level sets of conserved quantities $M^{s h}_{\mathfrak{c} m}$ are generically diffeomorphic to $2$-tori. 

We observed in \S 5.2 of \cite{G-V} that a generic trajectory on a 2-torus common level set $M_{\mathfrak{c} m}^{sh}$ is dense (see also Fig.~\ref{f:torus-plot-theta-phi}). This implies that any additional continuous conserved quantity would have to be constant everywhere on the torus and cannot be independent of the known ones. Thus, we may rule out additional independent conserved quantities.

\subsection{Classifying all common level sets of conserved quantities}
\label{s:common-level-set-conserved-qtys}

In \S \ref{s:Reduction-2D} we showed that the common level sets of the conserved quantities $\mathfrak{c}, m, s$ and $h$ are generically 2-tori. However, this leaves out some singular level sets. These non-generic common level sets occur when the conserved quantities fail to be independent and also correspond to the degeneration of the  elliptic function solutions (\ref{e:u-wp-function}) to hyperbolic and circular functions. Here,  we use a geometro-algebraic approach to classify all common level sets and show that there are only four possibilities: 2-tori, horn tori, circles and single points. Interestingly, the analysis relies on the properties of the cubic $\chi(u)$ that arose in the equation of motion for $u$ (\ref{e:EOM-u}).

\subsubsection{Common level sets as bundles and the cubic $\chi$}

We wish to identify the submanifolds of phase space $M^6_{S\text{-}L}$ obtained by successively assigning numerical values to the four conserved quantities $s, h, \mathfrak{c}$ and $m$. Not all real values of these conserved quantities lead to non-empty common level sets. From (\ref{e: H-mechanical}), we certainly need the Hamiltonian  $H \geq 0$ and $s^2 \geq 0$. It follows that $- s^2/2 - 1/2\la^2 \leq \mathfrak{c} \leq H/k^2 - 1/2\la^2$. However, these conditions are not always sufficient; additional conditions will be identified below. The situation is analogous to requiring the energy ($L_1^2 / 2 I_1 + L_2^2/ 2 I_2 + L_3^2 / 2 I_3$ in the principle axis frame) and square of angular momentum $(L_1^2 + L_2^2 + L_3^2)$ to be non negative for force-free motion of a rigid body. These two conditions are necessary but not sufficient to ensure that the angular momentum sphere and inertia ellipsoid intersect.
\begin{figure}[h]
	\centering
		\begin{subfigure}[t]{5cm}
		\centering
		\includegraphics[width=5cm]{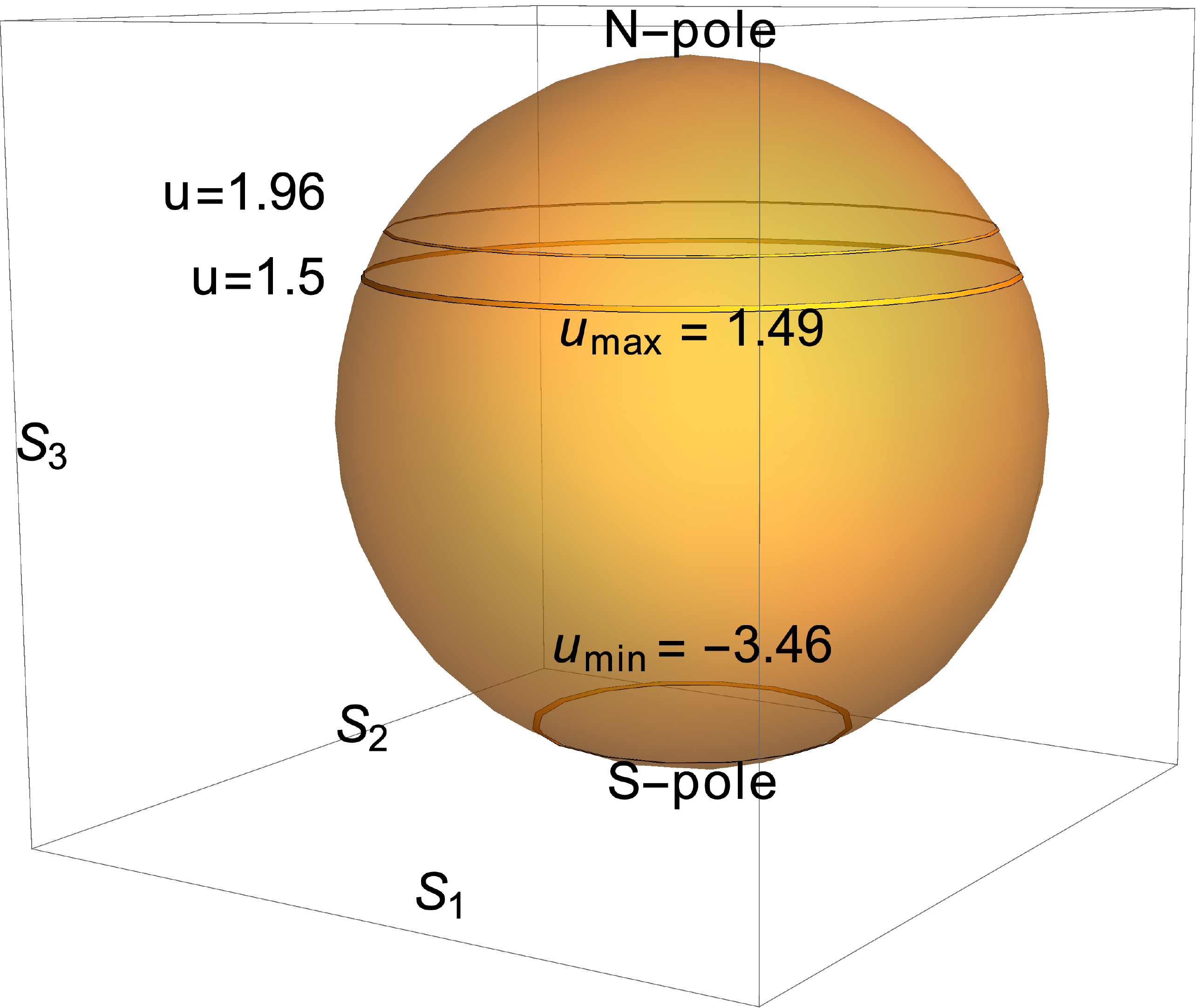}
		\caption{}
		\label{f:base-space}
		\end{subfigure}
		\qquad \quad
		\begin{subfigure}[t]{6cm}
		\centering
		\includegraphics[width= 6cm]{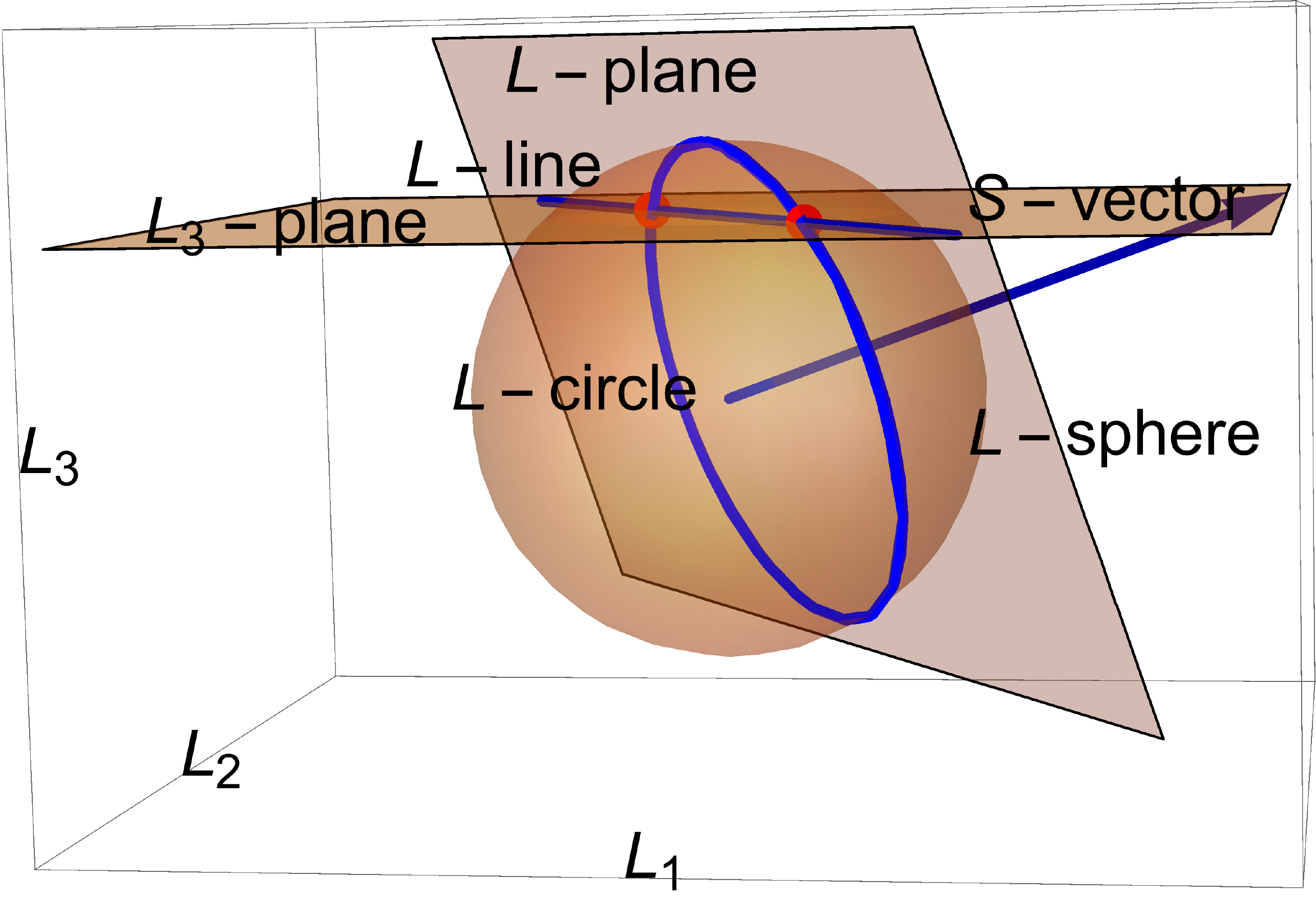}
		\caption{}
		\end{subfigure}
	\caption{ \footnotesize (a) The $S$-sphere $S_1^2 + S_2^2 + S_3^2 = 14 = s^2$ for $k = 1$. For $h = 1, \mathfrak{c} =2$ and $\la =1$, only latitudes below $u = S_3/k = 1.96$ (\ref{e:Intersection-L-plane-and-sphere}) are allowed if the $L$-sphere and $L$-plane are to intersect. However, if we take $m = -1$, the upper bound $u \leq (\la/2)(2 \mathfrak{c} - m^2)$ following from $L_1^2 + L_2^2 \geq 0$ and (\ref{e:conserved-quantities}) further restricts $u$ to lie below $1.5$. Finally, the condition $\chi \geq 0$ for non-empty fibres restricts $u$ to lie between the simple zeros $u_{\rm min} = -3.46$ and  $u_{\rm max} = 1.49$. (b) The $L$-space above the base point $\vec S = (3,2,1)$ for the same values of constants. The $L$-plane normal to $\vec S$ at a distance of $1/\sqrt{14}$ from $(0,0,0)$ is the level set $h = 1$. The $L$-sphere of radius $\sqrt{2}$ (the level set $\mathfrak{c}= 2$) intersects the $L$-plane along the $L$-circle. The horizontal $L_3$-plane $(L_3 = -m  = 1)$ intersects the $L$-plane along the $L$-line. The fibre over $\vec S$ is the pair of points where the $L$-line intersects the $L$-circle. The corresponding common level set is a 2-torus as in (C1) of \S \ref{s:Examples-of-CL-sets-of-conserved-quantities}.}
	\label{f:Common-level-set-L-space}
\end{figure}

First, putting $S_a S_a = s^2 k^2$ defines a $2$-sphere (the $`S$-sphere') in the $S$-space as in Fig.~\ref{f:base-space}. We may regard $u$ (or $S_3 = k u$) for $|u| \leq s$ as the latitude on the $S$-sphere with $u = \pm \sign{k}s$ representing the North $({\cal N})$ and South $({\cal S})$ poles. At each point on the $S$-sphere, the conservation of helicity $S_a L_a = k^2 h$ forces $\vec{L}$ to lie on a plane (the $`L$-plane') perpendicular to the numerical vector $\vec{S}$ at a distance $|h k|/s$ from the origin of the $L$-space. At this point, we have assigned numerical values to $s$ and $h$, which happen to be Casimirs of the Euclidean Poisson algebra (\ref{e:PB-SL-dual}). It remains to impose the conservation of $\mathfrak{c}$ and $m$.

For each point on the $S$-sphere, the condition $L_a^2/2 + k S_3/\la = \mathfrak{c} k^2$ (\ref{e:conserved-quantities}) defines an $L$-sphere of radius $\sqrt{2}|k|\left(\mathfrak{c} - u/\la\right)^\half  $in the $L$-space provided $\mathfrak{c} \geq u/\la$. Since $u \geq -s$, the conserved quantities must be chosen to satisfy $\mathfrak{c} \geq - s/\la$. In fact, this ensures that $H \geq 0$ and thus subsumes the latter. The $L$-sphere and the $L$-plane intersect along an $L$-circle provided the radius of the $L$-sphere exceeds the distance of the $L$-plane from the origin, i.e.,
	\beq
	|k|^{-1} \: {\rm rad}(L\text{-sphere}) = \sqrt{2\left(\mathfrak{c} - \frac{u}{\la} \right)} \geq \frac{ |h|}{s} = |k|^{-1} \: {\rm dist}(L\text{-plane}, {\bf 0})
	\quad \text{or} \quad  u \leq \la \left(\mathfrak{c} - \frac{h^2}{2s^2}\right).
	\label{e:Intersection-L-plane-and-sphere}
	\eeq
Thus, for the intersection to be non-empty, depending on the sign of $k$, $\vec S$ must lie below or above a particular latitude determined by (\ref{e:Intersection-L-plane-and-sphere}). Furthermore, since $u \geq -s$, we must choose 
	\beq
	\mathfrak{c} \geq \mathfrak{c}_{\rm min} = -  s/ \la + h^2 / 2 s^2.
	\label{e:C-lower-bound-strong}
	\eeq
When the inequality (\ref{e:Intersection-L-plane-and-sphere}) is saturated, the $L$-plane is tangent to the $L$-sphere and the $L$-circle shrinks to a point. In summary, the common level set of the three conserved quantities $s, h$ and $\mathfrak{c}$ can be viewed as a sort of fibre bundle with base given by the portion of the $S$-sphere lying above or below a given latitude. The fibres are given by $L$-circles of varying radii which shrink to a point along the extremal latitude.

The final conserved quantity $\Tr KL = mk^2$ restricts $\vec L$ to the horizontal plane $L_3 = -m k$. For each non-polar point on the $S$-sphere, this $L_3$-plane intersects the above $L$-plane along the $L$-line $S_1 L_1 + S_2 L_2 = hk^2 + mk S_3$ (assuming $S_1, S_2$ are not both zero). This line intersects the $L$-sphere at a pair of points, provided the radius of the $L$-sphere is greater than the distance of the $L$-line from the origin of the $L$-space, i.e.	
	\beq
	|k|^{-1} \: {\rm rad}(L\text{-sphere}) = \sqrt{2 \left(\mathfrak{c} - \frac{u}{\la}\right)} \geq \left( m^2 + \frac{(h + mu)^2 }{s^2 - u^2} \right)^{\half} 
	= |k|^{-1} \: {\rm dist}(L\text{-line}, {\bf 0}).
	\label{e:Intersection-L-line-and-L-sphere}
	\eeq
The two points of intersection coincide if the inequality is saturated so that the $L$-line is tangent to the $L$-sphere. Note that inequality (\ref{e:Intersection-L-line-and-L-sphere}) implies (\ref{e:Intersection-L-plane-and-sphere}), provided the $L$-sphere is non-empty ($\mathfrak{c} \geq u/\la $). This is geometrically evident since the distance of the $L$-line from the origin is bounded below by the distance $|k h|/s$ of the $L$-plane (which contains the $L$-line) from the origin.

\vspace{.25cm}

\footnotesize

{\fl \bf Remark:} Another way to see that (\ref{e:Intersection-L-line-and-L-sphere}) implies (\ref{e:Intersection-L-plane-and-sphere}) is to note that if $g =  m^2  + ((h + m u)^2/(s^2 - u^2)) - (h^2/s^2)$, then
	\beq
	\ov{k^2} \text{dist}(L\text{-line}, {\bf 0})^2 =  m^2  + \frac{(h + m u)^2}{s^2 - u^2} = \frac{h^2}{s^2} + g(u) = \ov{k^2} \text{dist}(L\text{-plane}, {\bf 0})^2 + g(u).
	\eeq	
Eq. (\ref{e:Intersection-L-line-and-L-sphere}) would then imply (\ref{e:Intersection-L-plane-and-sphere}), if we can show that $g(u) \geq 0$ on the sphere $|u| \leq s$. To see this, we first note that $g(u) \to + \infty$ at the poles $u = \pm s$ so that it suffices to show that the quadratic polynomial $\tilde{g}(u) =  g(u)(s^2 - u^2)$ is non-negative for $|u| < s$. This is indeed the case since the global minimum of $\tl g(u)$ attained at $u^* = -m s^2/h$ is simply zero.

\vspace{.25cm}

\normalsize
Assuming (\ref{e:Intersection-L-line-and-L-sphere}) holds, the common level set of all four conserved quantities may be viewed as a sort of fibre bundle with base given by the part of the $S$-sphere satisfying (\ref{e:Intersection-L-line-and-L-sphere}) and fibres given by either one or a pair of points (this is the case for non-polar latitudes, see below for the special circumstance that occurs above the poles). In other words, provided $\mathfrak{c} \geq \mathfrak{c}_{\rm min}$, the `base' space is the part of the $S$-sphere consisting of all latitudes $u$ lying in the interval $-s \leq u \leq \min(s, \la (\mathfrak{c} - h^2/2s^2))$ and satisfying the cubic inequality following from (\ref{e:Intersection-L-line-and-L-sphere})
	\beq
	\chi(u) = u^3 - \la \mathfrak{c} u^2 - \left( s^2 + \la h m \right) u + \frac{\la}{2}\left( 2 \mathfrak{c}s^2 - h^2 - m^2 s^2 \right) \geq 0.
	\label{e:cubic-equation-S3}
	\eeq
The roots of the cubic equation $\chi(u) = 0$ resulting from the saturation of this inequality determine the extremal latitudes where the two-point fibres degenerate to a single point (provided the extremal latitude does not correspond to a pole of the $S$-sphere). If an extremal latitude is at one of the poles then $\chi(\pm s) = -(\la/2)(h \pm ms)^2$ must vanish there and the determination of the fibre over the pole is treated below. 

Recall that the discriminant $\D = b^2 c^2 - 4 c^3 - 4 b^3 d - 27 d^2 + 18 bcd$ of the cubic $x^3 + b x^2 + c x + d$ is the product of squares of differences between its roots. It vanishes iff a pair of roots coincide. The discriminant of the cubic $\chi(u)$ will be useful in the analysis that follows. It is a function of the four conserved quantities: \small
	\beqs
	\D &=& \la^4 \mathfrak{c}^2 \left(\frac{s^2}{\la} +hm\right)^2 + 4 \la^3 \left(\frac{s^2}{\la} + hm\right)^3 + 2 \la^4 \mathfrak{c}^3(2\mathfrak{c} s^2 - h^2 - m^2 s^2)  -\frac{27}{4}\la^2 (2\mathfrak{c} s^2 - h^2 - m^2 s^2)^2 \cr
	&& + 9\la^3 \mathfrak{c}\left(\frac{s^2}{\la} + hm\right)(2 \mathfrak{c} s^2 - h^2 - m^2 s^2).
	\label{e:discriminant}
	\eeqs
	\normalsize	

\subsubsection{Fibres over the poles of the $S$-sphere}
\label{s:fibres-over-poles-of-S-sphere}

At the ${\cal N}$ and ${\cal S}$ poles $(u = \pm \sign{k}\, s)$ of the $S$-sphere, the $L$-plane $(S_3 L_3 = h k^2)$ and $L_3$-plane $(L_3 = -m k)$ are both horizontal: their intersection does not define an $L$-line. For the common level sets of $h$ and $L_3$ to be non-empty, the planes must coincide:
	\beq
	h = \mp m\, \sign{k} \: s
	\label{e:L-L3-plane}
	\eeq
with upper/lower signs corresponding to the ${\cal N}/{\cal S}$ poles. This condition ensures that $\chi$ vanishes at the corresponding pole, implying that it cannot be positive at a physically allowed pole of the $S$-sphere.

Now, for the $L$-sphere to intersect the $L_3$-plane, its radius must be bounded below by $|m k|$:
	\beq
	|k|^{-1} \: {\rm rad}(L\text{-sphere}) = \sqrt{2}\left( \mathfrak{c} \mp \frac{\sign{k}\, s}{\la}\right)^{\half}\geq |m| = |k|^{-1}\: {\rm dist}(L_3\text{-plane}, {\bf 0}).
	\label{e:condition-for-fibre-at-poles-of-S-sphere}
	\eeq
When this inequality is strict, the fibre over the pole is a circle ($L$-circle) while it is a single point when the inequality is saturated. Interestingly, in the latter case, the discriminant $\Delta$ (\ref{e:discriminant}) vanishes, so that the pole must either be a double or triple zero of $\chi$. On the other hand, when the inequality is strict, $\chi$ must have a simple zero at the pole. This structure of fibres over the poles is in contrast to the two point fibres over the non polar latitudes of the $S$-sphere when $\chi > 0$. For example, suppose $k = \la = s = 1$ and take $h = -m = 1$ so that the $L_3$ and $L$-planes over the ${\cal N}$ pole $(S_3 = 1)$ coincide. These planes intersect the $L$-sphere provided $\mathfrak{c} \geq 3/2$ (see (\ref{e:condition-for-fibre-at-poles-of-S-sphere})). Moreover, the fibre over the ${\cal N}$ pole is a single point if $\mathfrak{c} = 3/2$ and a circle if $\mathfrak{c} > 3/2$.   

\subsubsection{Properties of $\chi$ and the closed, connectedness of common level sets}

We observed in \S \ref{s:fibres-over-poles-of-S-sphere} that $\chi$ must vanish at a physically allowed pole of the $S$-sphere and that we must have $h = \pm m \: \sign{k} \: s$ for this to happen. Here, we investigate the possible behaviour of $\chi$ near a pole, which helps in restricting the allowed graphs of $\chi$. We find that the sign of $\chi'$ at  an allowed pole is fixed and also that the allowed latitudes must form a closed  and connected set. As a consequence, we deduce that some graphs of $\chi$ are disallowed. For example, $\chi$ cannot have a triple zero at a non-polar latitude. We also deduce that the common level sets must be both closed and connected. 


\vspace{.25cm}

{\fl \bf Result 1: Sign of $\chi'$ at a pole which is a {\it simple} zero:} Suppose $\chi$ has a simple zero  at the pole $u =  \pm s$ with non-empty fibre over it, then $\chi'(\pm s) \lessgtr 0$.
\vspace{.25cm}

\begin{figure}[h]
	\centering
		\begin{subfigure}[b]{13cm}
		\includegraphics[width=13cm]{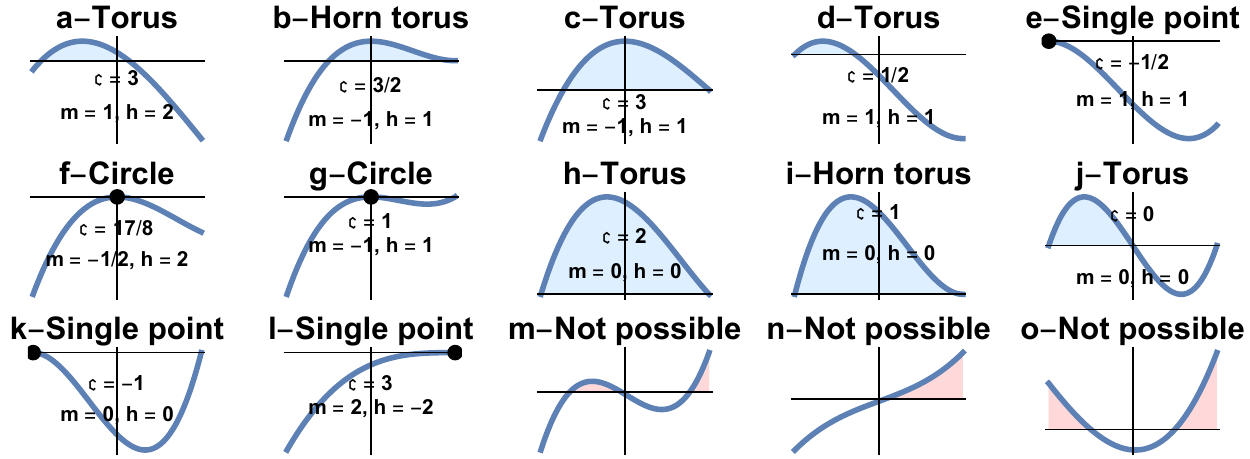}
		\end{subfigure}
	\caption{\footnotesize (a) - (l) Plots of the cubic $\chi(u)$ for latitudes between the south and north poles $- s \leq u \leq s$ for $k = \la = s = 1$ and $\mathfrak{c}, m$ and $h$ as indicated. The physically allowed latitudes with $\chi \geq 0$ are shaded in blue. The black dots indicate a single allowed latitude with $\chi$ necessarily having zeros of order more than one. The corresponding common level sets of conserved quantities (see \S \ref{s:Examples-of-CL-sets-of-conserved-quantities}) are a $2$-torus [(a), (c), (d), (h), (j)], a horn torus [(b), (i)], a circle [(f),(g)], and  a single point [(e), (k), (l)]. In (c), (d), (h) and (j) the fibre over the physically allowed poles (where $\chi$ has a simple zero) are circles while they are single points in (b), (e), (k) (double zero) and (l) (triple zero). In (i) the fibre over the ${\cal S}$ pole (simple zero) is a circle and is a point over the ${\cal N}$ pole (double zero). Similar figures with ${\cal N}$ and ${\cal S}$ exchanged arise when $k < 0$. Figures (m)-(o) show cases that {\it cannot} occur for any set of physically allowed conserved quantities as a consequence of {\bf Result} 2.}
	\label{f:Examples-of-common-level-set}
\end{figure}

{\fl \bf Proof of $\chi'(s) < 0$:} Suppose $h = -m s$, so that $\chi(u)$ has a simple zero at the pole $u = s$ with circular fibre over it (see Eq.(\ref{e:L-L3-plane})). Then (\ref{e:cubic-equation-S3}) implies
	\beq
	\chi'(s) = 2 s^2 - \la s (2 \mathfrak{c} - m^2).
	\eeq
Suppose $\chi'(s) > 0$, then $\mathfrak{c} < s/ \la + m^2/2$. But in this case, the upper bound on the latitude $u \leq \min[s, \la \mathfrak{c} - \la h^2/ (2 s^2)] < s$ so that $u = s$ could not have been an allowed latitude. On the other hand, if $\chi'(s) < 0$, then $u=s$ is an allowed latitude. Thus, when the ${\cal N}/{\cal S}$ pole for $k \gtrless 0$ is a simple zero of $\chi$ with non-empty fibre, it is always surrounded by other allowed latitudes. In particular, the north poles in Fig.~\ref{f:Examples-of-common-level-set}g, j and k are not allowed latitudes, while they {\it are} in Fig.~\ref{f:Examples-of-common-level-set}c and h.
\vspace{.25cm}

{\fl \bf Proof of $\chi'(-s) > 0$:} On the other hand, suppose $h = m s$ so that $\chi$ has a simple zero at $u = -s$ with non-empty fibre. Suppose $\chi'(-s) < 0$, then as before (\ref{e:cubic-equation-S3}) implies $\mathfrak{c} < -s/ \la + m^2/2 \leq \mathfrak{c}_{\rm min}$ which violates (\ref{e:C-lower-bound-strong}). Thus $\chi'(-s)$ must be positive. In other words, when the pole $u= -s$ is a simple zero of $\chi$ with non-empty fibre, it must be surrounded by other allowed latitudes. So the poles cannot be simple zeros unless the neighbouring latitudes are allowed. In particular, the south poles in Fig.~\ref{f:Examples-of-common-level-set}d, h, i and j are allowed latitudes.

\vspace{.25cm}

{\bf \fl Result 2: Set of allowed latitudes and common level set must be closed:} The conserved quantities $\mathfrak{c}, m, s$ and $h$ define continuous functions (quadratic in $S$ and $L$) from the phase space $M^6_{S\text{-}L}$ to the four-dimensional space $\cal Q$ of conserved quantities (which is a subset of $\mathbb{R}^4$ consisting of the 4-tuples $(\mathfrak{c}, m ,h, s)$ subject to the conditions $s \geq 0$ and $\mathfrak{c} \geq \mathfrak{c}_{\rm min}$ (\ref{e:C-lower-bound-strong})). Each of their common level sets must be a closed subset of $M^6_{S\text{-}L}$ as it is the inverse image of a point in $\cal Q$. We may use this to deduce that $\chi$ cannot approach a positive value at a pole. We have already observed that if a pole is an allowed latitude then $\chi$ must vanish there. On the other hand, suppose a pole $P$ is not an allowed latitude but $\chi$ is positive in a neighbourhood of $P$. Then the set of allowed latitudes would be an open set and so would the common level set. In particular, $\chi$ cannot have (i) only one simple zero on the $S$-sphere and be non-vanishing elsewhere (as in Fig.~\ref{f:Examples-of-common-level-set}n) (ii) three simple zeros between the poles (see Fig.~\ref{f:Examples-of-common-level-set}m) (iii) a double zero and a simple zero between the poles (iv) a triple zero at a non-polar latitude (v) two simple zeros between the poles with $\chi > 0$ at the poles (as in Fig.~\ref{f:Examples-of-common-level-set}o) or (vi) a double zero between the poles with $\chi > 0$ at the poles.

\vspace{.25cm}

{\fl \bf Common level set of conserved quantities must be connected:} For the common level set to be disconnected, the set of allowed latitudes on the $S$-sphere must be disconnected. The only remaining way that this could happen is for $\chi$ to have three distinct simple zeros on latitudes $u \in [- s, s]$ of the $S$-sphere. Let us show that this is disallowed. Now {\bf Result} 2 prevents $\chi$ from having three simple zeros at non-polar latitudes. It only remains to consider the cases where either of the poles is a simple zero of $\chi$. If $\chi$ has a simple zero at $s$, then by {\bf Result} 1, $\chi'(s) < 0$. Since $\chi(\infty) = \infty$, $\chi$ can have at most one more zero on the $S$-sphere so that the set of allowed latitudes is connected. On the other hand, suppose $\chi$ has a simple zero at $-s$, then $\chi'(-s) > 0$ by {\bf Result} 1. Suppose further that $\chi$ has two more simple zeros $-s < u^* < u^{**} \leq s$ on the $S$-sphere, then by {\bf Result} 2, $u^{**}$ must equal $s$ as otherwise $\chi$ would be positive at the pole $u = s$ as in the disallowed Figs.~\ref{f:Examples-of-common-level-set}m, n and o. So $u^{**} = s$ with $\chi'(s) > 0$ as in Fig.~\ref{f:Examples-of-common-level-set}j. But in this case, {\bf Result} 1 forbids $u^{**}$ from being an allowed latitude, so that the set of allowed latitudes is again a single interval $[-s, u^*]$.

\vspace{.25cm}

{\fl \bf Triple zeros of $\chi$:} For $\chi(u)$ (\ref{e:cubic-equation-S3}) to have a triple zero, i.e., to be of the form $(u - z)^3$, we must have $z = \la \mathfrak{c}/3$ and the conserved quantities must satisfy two conditions:
	\beq
	\mathfrak{c}^2 = -\frac{3}{\la} \left(hm +\frac{s^2}{\la}\right) \quad \text{and} \quad 2 \la^2 \mathfrak{c}^3= -27(2\mathfrak{c}s^2 - h^2 - m^2s^2).
	\label{e:chi-triple-0-condition-on-cons-qty}
	\eeq
These conditions define a two-dimensional surface in the space $\cal Q$ of conserved quantities. {\bf Result} 2 implies that $\chi$ cannot have a triple zero at a non-polar latitude. On the other hand, $\chi$ {\it can} have a triple zero at ${\cal N}$ or ${\cal S}$ provided both (\ref{e:L-L3-plane}) and (\ref{e:chi-triple-0-condition-on-cons-qty}) are satisfied. Putting $h = \mp \sign{k} \, ms$ in (\ref{e:chi-triple-0-condition-on-cons-qty}), the conditions for $\cal N$ or $\cal S$ to be a triple zero become 
	\beq
	\pm 3 \la \, \sign{k} \, s m^2 = \la^2 \mathfrak{c}^2 + 3 s^2 \quad \text{and} \quad  
	\la \mathfrak{c} = 3 s.
	\label{e:triple-zero-condition-chi}
	\eeq
The first condition implies that $\chi$ cannot have a triple zero at ${\cal S}$ for $k > 0$ or at ${\cal N}$ for $k < 0$. On the other hand, $\chi$ {\it can} have a triple zero at $\cal N$ for $k > 0$ as in Fig.~\ref{f:Examples-of-common-level-set}l.

\subsubsection{Possible types of common level sets of all four conserved quantities}
\label{s:Examples-of-CL-sets-of-conserved-quantities}

Here we combine the above results on the connectedness of common level sets, slope of $\chi$ at the poles and on the structure of the fibres over polar and non-polar latitudes of the $S$-sphere to identify all possible common level sets of conserved quantities. There are only four possibilities: the degenerate or singular level sets (horn tori, circles and single points) and the generic common level sets (2-tori). These possibilities are distinguished by the location of roots of $\chi$. They are discussed below and illustrated in Fig \ref{f:Examples-of-common-level-set}. In (C1)-(C5) below we take $k > 0$ so that $u = \pm s$ correspond to the $\cal N$ and $\cal S$ poles. Similar results hold for $k < 0$ with $\cal N$ and $\cal S$ interchanged.



(C1) For generic values of conserved quantities, $\chi(u)$ is positive between two neighbouring non-polar simple zeros $u_{\rm min} < u_{\rm max}$ lying in $(-s, s)$ (E.g. $k = \la = s = m = 1$, $h = 2$ and $\mathfrak{c} = 3$ as in Fig.~\ref{f:Examples-of-common-level-set}a). The base space of \S\ref{s:common-level-set-conserved-qtys} is the portion of the $S$-sphere lying between the latitudes $u_{\rm min}$ and $u_{\rm max}$, with the two-point fibres shrinking to single point fibres along the extremal latitudes $u_{\rm min}$ and $u_{\rm max}$. The resulting common level set is homeomorphic to a pair of finite coaxial cylinders with top as well as bottom edges identified, i.e., a $2$-torus.


To visualize the above toroidal common level sets and some of its limiting cases which follow, it helps to qualitatively relate the separation between zeros of $\chi$ to the geometric parameters of the torus embedded in three dimensions. For instance, the minor diameter of the torus grows with the distance between $u_{\rm min}$ and $u_{\rm max}$. Thus, when the simple zeros coalesce at a double zero, the minor diameter vanishes and the torus shrinks to a circle. Similarly (for $k > 0$) the major diameter of the torus grows with the distance between $u_{\rm min}$ and $\cal N$. Thus, when $u_{\rm max} \to {\cal N}$, the major and minor diameters become equal and we expect the torus to become a horn torus. However, this requires the fibre over ${\cal N}$ to be a single point, which is true only when $\cal N$ is a double zero of $\chi$. 


(C2) A limit of (C1) where either $u_{\rm min} \to {\cal S}$ or $u_{\rm max} \to {\cal N}$ and $\chi$ is positive between them. For instance, if $u_{\rm max} \to {\cal N}$ and the fibre over $\cal N$ is a single point, then the common level set is homeomorphic to a horn torus (E.g. $\la = k = s= h = 1$, $m =-1$ and $\mathfrak{c} = 3/2$ as in Fig.~\ref{f:Examples-of-common-level-set}b). On the other hand, for $\mathfrak{c} > 3/2$ the fibre over ${\cal N}$ is a circle and we expect the common level set to be a $2$-torus (see Fig.~\ref{f:Examples-of-common-level-set}c). It is as if the circular fibre over the single-point latitude $\cal N$ plays the role of an extremal circular latitude with single point fibre in (C1), thus the roles of base and fibre are reversed. Similarly, when $u_{\rm min} \to {\cal S}$ with circular fibre over ${\cal S}$, the common level set is homeomorphic to a 2-torus (E.g. $k = \la = s = m = h = 1$ and $\mathfrak{c} > -1/2$ as in Fig.~\ref{f:Examples-of-common-level-set}d). In the limiting case where $\mathfrak{c} = \mathfrak{c}_{\rm min} = -1/2$, the two simple zeros $u_{\rm min}$ and $u_{\rm max}$ merge at $\cal S$. The fibre over $\cal S$ becomes a single point and the common level shrinks to a point (see Fig.~\ref{f:Examples-of-common-level-set}e).
 

(C3) Another limit of (C1) where the roots $u_{\rm min}$ and $u_{\rm max}$ coalesce at a double root $u_d \in (-s, s)$ of $\chi$. $\chi$ is negative on the $S$-sphere except along the latitude $u_d$ and the fibre over it is a single point. The discriminant $\D$ (\ref{e:discriminant}) must vanish for this to happen. The common level set becomes a circle corresponding to the latitude $u_d$. For example, if $k = \la =1$ and $s = 1, m = -1/2, h = 2$ and $\mathfrak{c} = 17/8$, then the equator $u_d = 0$ is the allowed latitude as shown in Fig.~\ref{f:Examples-of-common-level-set}f. Another example of a circular common level set appears in Fig.~\ref{f:Examples-of-common-level-set}g. In this case {\bf Results} 1 and 2  exclude the north pole ensuring the connectedness of the common level set.


(C4) A limit of (C1) where the simple zeros $u_{\rm min}$ and $u_{\rm max}$ move to ${\cal S}$ and ${\cal N}$ respectively, with $\chi > 0$ in between. In this case, both poles have circular fibres and the common level set is a 2-torus. This happens, for instance, when $\mathfrak{c} \to \infty$, irrespective of the values of $m, h$ and $s > 0$. Another way for this to happen is for $m$ and $h$ to vanish so that the poles are automatically zeros of
	\beq
 	\chi(u) = u^3 - \la \mathfrak{c} u^2 - s^2u + \la \mathfrak{c} s^2 = (u - s)(u + s)(u - \la \mathfrak{c}) \quad [\text{for} \quad m = h = 0]
	\eeq
and to choose $\mathfrak{c} > s/\la$ to ensure there is no zero in between. Holding $s, h$ and $m$ fixed, three more possibilities arise as we decrease $\mathfrak{c}$. When $\mathfrak{c} = s/\la$, $\chi$ has a double zero at ${\cal N}$ (Fig.~\ref{f:Examples-of-common-level-set}i) with a single point fibre over it and the common level set becomes a horn torus.  For $-s/\la < \mathfrak{c} < s/\la$, the third zero of $\chi$ moves from $\cal N$ to the latitude  $u = \la \mathfrak{c}$. By {\bf Result} 1, the allowed latitudes go from $u= -s$ to $u = \la \mathfrak{c}$ (see Fig.~\ref{f:Examples-of-common-level-set}j), and the common level set returns to being a  $2$-torus. Finally, when $\mathfrak{c} = \mathfrak{c}_{\rm min} = -s/\la$, the only allowed latitude $(\cal S)$ is a double zero and the common level set shrinks to a point (see Fig.~\ref{f:Examples-of-common-level-set}k).   


(C5) $\chi$ has a zero at just one of the  poles and is negative elsewhere on the $S$-sphere. The common level set is then a single point. We encountered this as a limiting case of (C2) where $\chi$ has a double zero at $\cal S$ as in Fig.~\ref{f:Examples-of-common-level-set}e. This can also happen when $\chi$ is negative on the $S$-sphere except for a triple zero at either ${\cal S}$ ($k < 0$) or ${\cal N}$ ($k > 0$) (see Eq. (\ref{e:triple-zero-condition-chi})). For example, when $k= \la = s = 1$, $\mathfrak{c} = 3, m = 2$ and $h=-2$, $\chi$ has a triple zero at ${\cal N}$ as in Fig.~\ref{f:Examples-of-common-level-set}l.

\subsection{Nature of the `Hill' region and energy level sets using Morse theory}
\label{s:Hill-region-Morse-theory}

In this section, we study the `Hill' region $W^{E}_{\mathfrak{c} m}$, which we define as the set of points on the symplectic leaf $M^4_{\mathfrak{c} m}$ with energy less than or equal to $E$:	
	\beq
	W^{E}_{\mathfrak{c} m} = \{ p \in M^4_{\mathfrak{c} m} | H(p) \leq E \}.
	\eeq
The $H = E k^2$ energy level set $M^E_{\mathfrak{c} m}$ is then the boundary of $W^{E}_{\mathfrak{c} m}$. Taking $R_{1,2}$ and $P_{1,2}$ (\ref{e:R-P-relation-to-L-S}) as coordinates on $M^4_{\mathfrak{c} m}$, we treat the Hamiltonian
	\beq
	\frac{H}{k^2} = \frac{P_1^2 + P_2^2}{2} + \frac{\la m}{2}(R_1 P_2 -R_2 P_1) + \frac{\la^2}{8} (R_1^2 + R_2^2) \left(R_1^2 + R_2^2 + 3 m^2 - 4 \mathfrak{c} \right) + \frac{\la^2}{8} (2 \mathfrak{c} - m^2)^2 + \mathfrak{c} + \frac{1}{2 \la^2}
	\eeq
as a Morse function \cite{Milnor}. The nature of critical points of $H$ depends on the value of $2 \mathfrak{c} - m^2$. There are two types of critical points: (a) an isolated critical point at $R_{1,2} = P_{1,2} = 0$ which exists for all values of $2 \mathfrak{c} - m^2$ and (b) a ring of critical points 
	\beq
	R_1^2 + R_2^2 = 2 \mathfrak{c} - m^2 \quad \text{with} \quad ( P_1, P_2 ) = \frac{\la m}{2} \left( R_2, -R_1 \right),
	\eeq
which exists only for $2 \mathfrak{c} - m^2 > 0$ and shrinks to the isolated critical point when $2 \mathfrak{c} - m^2  = 0$. The energy at these critical points is
	\beq
	E_{\rm iso} = \frac{\la^2}{8} (2 \mathfrak{c} - m^2)^2 + \mathfrak{c} + \frac{1}{2 \la^2} \quad \text{and} \quad
	E_{\rm ring} =  \mathfrak{c} + \ov{2 \la^2}.
	\eeq
Upon varying $\mathfrak{c}$ and $m$, the isolated critical points cover all of the static submanifold $\Sigma_2$ while the rings of critical points cover the static submanifold $\Sigma_3$. By finding the eigenvalues of the Hessian of the Hamiltonian at these critical points, we find that for $2 \mathfrak{c} - m^2 < 0$ the isolated critical point G is a local minimum of energy (four +ve eigenvalues). In fact, for $2 \mathfrak{c} - m^2 < 0$, the isolated critical point has to be the global minimum of energy as the energy is bounded below and there are no other extrema of energy. For $2 \mathfrak{c} - m^2 > 0$, the isolated critical point becomes a saddle point (two +ve and two -ve eigenvalues) with energy $E_{\rm sad} = E_{\rm iso}$. On the other hand, the ring of critical points are degenerate global minima (three +ve and one zero eigenvalue). To apply Morse theory, we need the indices of the critical points of $H$ (number of negative eigenvalues of the Hessian). From the foregoing, we see that the ground state G has index zero, the saddle point has index two and the degenerate critical points on the ring may be nominally assigned a vanishing index.

\vspace{.25cm}

{\fl \bf Change in topology of the Hill region:} According to Morse theory \cite{Milnor}, the topology of the Hill region can change only at critical points of the Hamiltonian. (a) For $2 \mathfrak{c} - m^2 < 0$, there is only one critical point, the global minimum G with index zero and energy $E_{\rm G} = E_{\rm iso}$. Thus, as $E$ increases beyond $E_{\rm G}$, the Hill region $W^E_{\mathfrak{c} m}$ goes from being empty to being homeomorphic to a 4-ball $(B^4 = \{ {\bf x} \in \mathbb{R}^5 \quad \text{with} \quad  \Vert {\bf x} \Vert \leq 1 \})$ arising from the addition of a 0-cell. (b) For $2 \mathfrak{c} - m^2 > 0$, there are two critical values of energy $E_{\rm ring} < E_{\rm sad}$ corresponding to the ring of critical points and the saddle point. The index vanishes along the ring of critical points, so when $E$ crosses $E_{\rm ring}$, the Hill region acquires a 3-ball (0-cell) for each point on the ring corresponding to the 3 positive eigenvalues of the Hessian. Thus $W^{E}_{\mathfrak{c} m}  \cong  B^3 \times S^1$ for $E_{\rm ring} < E < E_{\rm sad}$. The saddle point with $E = E_{\rm sad}$ has index two, so the topology of $W^{E}_{\mathfrak{c} m}$ changes to $B^4$ upon adding a 2-cell to $B^3 \times S^1$ (the analogous statement in one lower dimension is that adding a 2-cell to the hole of the solid torus $(B^2 \times S^1)$ gives a $B^3$).

\vspace{.25cm}

{\fl \bf Nature of energy level sets:} The energy level set $M^E_{\mathfrak{c} m}$ is the boundary of the Hill region, i.e. $M^E_{\mathfrak{c} m} = \pdr W^E_{\mathfrak{c} m}$. It is a 3-manifold except possibly at the critical energies. Thus for $2\mathfrak{c} - m^2 < 0$, $M^E_{\mathfrak{c} m} \cong \pdr B^4 \cong S^3$ for all energies $E > E_{\rm G}$. On the other hand, when $2\mathfrak{c} - m^2 > 0$ the energy level set undergoes a change in topology from $S^2 \times S^1$ to $S^3$ as $E$ crosses $E_{\rm sad}$. 

The energy level sets at the critical values $E_{\rm G}, E_{\rm sad}$ and $E_{\rm ring}$ are exceptional. For given $\mathfrak{c}$ and $m$ with $2 \mathfrak{c} - m^2 < 0$ and $E = E_{\rm G}$, $M^E_{\mathfrak{c} m}$ is a single point on $\Sigma_2$ (the critical point), since G is the non-degenerate global minimum of energy. When $2 \mathfrak{c} - m^2 > 0$, $E = E_{\rm sad}$ fixes $s = (\la/2)(2 \mathfrak{c} - m^2)$ leaving a range of possible values of $h \in (h_{\rm min} , h_{\rm max})$, whose values are determined by eliminating $u$ from the conditions $\chi(u) = \chi'(u) = 0$. This leads to a three-dimensional energy level set. $M^{E_{\rm sad}}_{\mathfrak{c} m}$ includes one horn torus with its center as the saddle point for $h = h_{\rm sad}$ as well as a one parameter family of toroidal level sets for $h_{\rm min} < h \neq h_{\rm sad}  < h_{\rm max}$ and a pair of  circular level sets occurring at $h_{\rm min}$ and $h_{\rm max}$. Interestingly, horn tori arise only  when $E= E_{\rm sad}$, since $s = (\la/2)(2 \mathfrak{c} - m^2)$ is a necessary condition for horn tori (see \S \ref{s:horn-toroidal-level-sets}). Thus, the horn torus is a bit like the figure-8 shaped separatrix one encounters in particle motion in a double well potential. Finally, the $E = E_{\rm ring}$ level manifold consists of a ring of single point common level sets, each lying on the static submanifold $\Sigma_3$. Unlike static solutions and horn tori, circular and 2-toroidal level sets also arise at non-critical energies.

\section{Foliation of phase space by tori, horn tori, circles and points}
\label{s:foliation-of-phase-space}

For generic allowed values of the conserved quantities $\mathfrak{c}, m, s$ and $h$, their common level set in the $M^6_{S \text{-} L}$ phase space is a $2$-torus. As noted, this happens when $\chi$ has simple zeros along a pair of latitudes of the $S$-sphere and is positive between them. However, this $4$-parameter family of invariant tori does not completely foliate the phase space: there are some other `singular' level sets as well: horn tori, circles and points. The union of single-point level sets is $\Sigma_2 \cup \Sigma_3$ (\ref{e:static-submanifolds}), consisting of static solutions. They occur when $\chi(u)$ has a triple zero at $u = s$ or is a local maximum at a double zero at $u = \pm s$. We will now discuss the other cases in increasing order of complexity. In each case, we view the union of common level sets of a given type as the state space of a self-contained dynamical system which has the structure of a fibre bundle over an appropriate submanifold of the space $\cal Q$ of conserved quantities. The fibres in each case are circles, horn tori and tori. The dynamics on the union of circles and tori is Hamiltonian and we identify action-angle variables on them. On the other hand, we show that the dynamics on the union of horn tori is a gradient flow.

\subsection{Union $\cal C$ of circular level sets: Poisson structure \& action-angle variables}
\label{s:circular-level-sets}

In this section, we show that the union of circular level sets is the same as the trigonometric/circular submanifold $\cal C$ (introduced in \S 5.6 of \cite{G-V}) where the solutions are sinusoidal functions of time. Local coordinates on $\cal C$ are furnished by $\mathfrak{c}, m, u$ and $\tht$ (or equivalently $\phi$) and we express the Hamiltonian in terms of them. The Poisson structure on $\cal C$ is degenerate with $\mathfrak{c}$ and $m$ generating the center and their common level sets being the symplectic leaves. While $u$ is a constant of motion, $\tht$ evolves linearly in time. We exploit these features to obtain a set of action-angle variables for the dynamics on $\cal C$.

\subsubsection{$\cal C$ as a circle bundle and dynamics on it}

As pointed out in example (C3) of \S\ref{s:Examples-of-CL-sets-of-conserved-quantities}, the common level set of conserved quantities is a circle when the cubic $\chi(u)$ (\ref{e:cubic-equation-S3}) has a double zero at a non-polar latitude of the $S$-sphere and is negative on either side of it. In this case, the latitude $u$ is restricted to the location of the double zero. To identify the three-dimensional hypersurface ${\cal Q}_{\cal C}$ in the four-dimensional space $\cal Q$ of conserved quantities, where $\chi$ has a double zero at a non-polar latitude, we will proceed in two steps.  First, we compare the equation $\chi = 0$ with $(u - u_2)^2(u - u_1) = 0$ to arrive at the three conditions:
	\beq
	2 u_2 + u_1 = \la  \mathfrak{c}, \quad
	u_2^2 + 2 u_2 u_1 = -\left( s^2 + h m \la \right) \quad \text{and} \quad
	-u_2^2 u_1 = \frac{\la}{2} \left((2\mathfrak{c} - m^2) s^2 - h^2 \right).
	\label{e:relations-for-chi-double-zero}
	\eeq
The first two may be used to express the roots $u_2$ and $u_1$ in terms of conserved quantities:
	\beq
	u_2^\pm = (1/3) \left( \la \mathfrak{c}  \pm \sqrt{\la^2 \mathfrak{c}^2 + 3( s^2 + \la h m)}\right) \quad \text{and} \quad 
	u_1^\pm = \la \mathfrak{c} - 2u_2.
	\label{e:roots-u1-u2}
	\eeq
The third equation in (\ref{e:relations-for-chi-double-zero}) then leads to the following conditions among conserved quantities
	\beq
	27 \la h^2 - 36 \la \mathfrak{c} s^2  + 27 \la m^2 s^2  + 18 \la^2 \mathfrak{c} h m  + 4  \la^3 \mathfrak{c}^3  = \mp 4(3 s^2 + \la (3 h m + \la \mathfrak{c}^2 ))^{3/2}.
	\label{e:relations-circle-level-set}
	\eeq
Squaring, these conditions are equivalent to $\Delta = 0$, where $\D$ is the discriminant (\ref{e:discriminant}) of $\chi$. The three-dimensional submanifold of $\cal Q$ defined by $\D = 0$, however, includes 4-tuples ($\mathfrak{c}$, $m$, $s$, $h$) corresponding to horn toroidal (double zero at the pole $u=s$) or single-point (triple zero at $u = s$ or double zero at $u= s$ or $-s$) common level sets, in addition to circular level sets. To eliminate the former, we must impose the further conditions $u_2 \neq u_1$, $|u_2| < s$ and $\chi''(u_2) < 0$. This last condition, which says $u_2 <\la \mathfrak{c}/3$, selects the roots $u_{1,2} = u_{1,2}^-$ in (\ref{e:roots-u1-u2}). These conditions define the three-dimensional hypersurface ${\cal Q}_{\cal C} \subset \cal Q$ corresponding to circular level sets. Now, $\mathfrak{c}, m$ and $s$ may be chosen as coordinates on ${\cal Q}_{\cal C}$, with (\ref{e:relations-circle-level-set}) allowing us to express $h$ in terms of them. Interestingly, we find by studying examples, that  for values of $\mathfrak{c}, m$ and $s$ corresponding to a circular level set, there are generically two distinct values of $h$; so we would need two such coordinate patches to cover ${\cal Q}_{\cal C}$. The union of all these circular level sets may be viewed as a sort of circle bundle over ${\cal Q}_{\cal C}$ and forms a four-dimensional `circular' submanifold $\cal C$ of $M^6_{S\text{-}L}$. As shown in \S 5.5 and \S 5.6 of \cite{G-V}, this circular submanifold along with its boundary coincides with the set where the four-fold wedge product $dh \wedge ds^2 \wedge dm \wedge d\mathfrak{c}$ vanishes. 

The equations of motion (\ref{e: EOM-LS}) simplify on the circular submanifold $\cal C$. Indeed, since $S_3 = ku$ is a constant, $\dot S_3 = 0$ so that $S_1/S_2 = L_1/L_2$ implying that $\tht - \phi = n \pi$ where $n \in {\mathbb Z}$. As shown in \S 5.5 of \cite{G-V}, the equations of motion then simplify to 
	\beq
	\dot{S_1} = -\dot \phi S_2, \qquad
	\dot{S_2} = \dot\phi S_1, \qquad
	\dot{L_1} = k S_2 \qquad \text{and} \qquad
	\dot{L_2} = -k S_1
	\eeq 
with sinusoidal solutions:
	\beq
	S_1/k = A \sin k \om t + B \cos k \om t \quad \text{and} \quad 
	S_2/k = A \cos k \om t - B \sin k \om t.
	\label{e:circular-level-set-solutions}
	\eeq
Here, using (\ref{e:L-S-polar}), $\omega = S_{1,2}/L_{1,2} = (-1)^{n} \rho/r = -\dot \phi/k = -\dot \tht/k$, which varies with location on the base ${\cal Q}_{\cal C}$. It is the non-dimensional angular velocity for motion in the circular fibres. Since $\rho$ and $r$ are positive, $(-1)^n \omega = |\omega|$. Here both $\tht$ and $\phi$ evolve linearly in time and the equality of 
	\beq
	\dot \tht = (-1)^{n+1} \frac{k \rho}{r} \quad \text{and} \quad \dot \phi = k \la \left(m + (-1)^n \frac{u r}{\rho} \right)
	\label{e:circular-levelset-dynamics}
	\eeq 
implies that the constant of motion $u$ may be expressed in terms of $\omega$ and $m$: 
	\beq
	u = - \om (m + \om/\la).
	\label{e:quadratic-alpha}
	\eeq
\footnotesize
{\fl \bf Remark:} If the $S$-sphere shrinks to a point $(s = h = 0)$ then one still has circular level sets consisting of latitudes of the $L$-sphere determined by $m$, provided $2\mathfrak{c} \geq m^2$. However, each point on these exceptional circular level sets is a static solution lying on $\Sigma_3$ (\ref{e:static-submanifolds}).
\normalsize
	
\subsubsection{Canonical coordinates on $\cal C$}
{\fl \bf Local coordinates on $\cal C$:} For the analysis that follows, a convenient set of coordinates on the `circle bundle' $\cal C$ consists of $\mathfrak{c}, m$ and $\omega$ for the base ${\cal Q}_{\cal C}$ and $\tht$ for the fibres. The dynamics on $\cal C$ admits three independent conserved quantities as there is one relation among $\mathfrak{c}, m , s$ and $h$ following from (\ref{e:relations-circle-level-set}). Since the common level sets of the conserved quantities on $\cal C$ are circles, rather than tori, it is reasonable to expect there to be two Casimirs (say $\mathfrak{c}$ and $m$) for the Poisson structure on $\cal C$, as we show below. In fact, $\cal C$ is foliated by the common level surfaces of $\mathfrak{c}$ and $m$ (symplectic leaves) which serve as phase spaces (with coordinates $\omega$ and $\tht$) for a system with one degree of freedom. $\tht$  is then the coordinate along the circular level sets of the Hamiltonian on these two-dimensional symplectic leaves. 

To find the reduced Hamiltonian on $\cal C$ we express the remaining variables in terms of $\mathfrak{c}, m, \omega$ and $\tht$. The formula for $\mathfrak{c}$ (\ref{e:conserved-quantities}) along with (\ref{e:quadratic-alpha}) determines $r^2 \equiv 2\mathfrak{c} - m^2 + (2 \om/\la) (m + \om/\la)$ and consequently $\rho = |\omega| r$ as well. The remaining conserved quantities are given by
	\beqs
	h &=& (-1)^n \rho r - m u = \omega \left( 2\mathfrak{c} - m^2 + \frac{2 \om}{\la}\left( m + \frac{\om}{\la} \right)  \right) + m \om \left(m + \frac{\om}{\la} \right)  \quad \text{and} \cr
	s^2 &=& \rho^2 + u^2 = 2 \om^2 \left(\mathfrak{c}  + \frac{2 m \om}{\la} + \frac{3 \omega^2}{2 \la^2}\right). 
	\label{e:h-and-s-on-circular-LS}
	\eeqs
Thus, the reduction of the Hamiltonian (\ref{e: H-mechanical}) to the trigonometric submanifold is
	\beq
	H(\mathfrak{c}, m, \omega) = k^2  \left( \om^2 \left(\mathfrak{c} + \frac{2 m \omega}{\la} + \frac{3 \omega^2}{2 \la^2}  \right) + \mathfrak{c} + \frac{1}{2 \la^2} \right).
	\label{e:H-trigonometric}
	\eeq
As remarked, for given values of $\mathfrak{c}, m$ and $s$, there are generically two possible values of $h$ corresponding to two points on ${\cal Q}_{\cal C}$. By considering examples, we verified that for each of them, there is a unique $\om$ that satisfies (\ref{e:quadratic-alpha}) and both the equations in (\ref{e:h-and-s-on-circular-LS}).

\vspace{.25 cm}

{\fl \bf Poisson structure on $\cal C$:} We wish to identify Poisson brackets among the coordinates $\mathfrak{c}, m, \omega$ and $\tht$ that along with the reduced Hamiltonian (\ref{e:H-trigonometric}) gives the equation of motion $\dot{\tht}= -\om k$ on $\cal C$. As noted, it is natural to take $\mathfrak{c}$ and $m$ as Casimirs so that $\{ \mathfrak{c} , m\} = \{ \mathfrak{c} , \om \} = \{m, \om \} = \{ \mathfrak{c}, \tht \} = \{ m , \tht \} = 0$. The only non-trivial Poisson bracket $\{ \tht, \omega \}$ is then determined as follows from (\ref{e:H-trigonometric}):
	\beq
	\dot \tht = - k \omega = \{ \tht, H \} = \pdr_\om H \{ \tht, \omega \} \quad \imply \quad
	\{ \tht, \omega \} = - \frac{k \omega}{\pdr_\om H} = -\frac{1}{2 k} \left(\mathfrak{c} + \frac{3 \om}{\la}\left( m + \frac{\om}{\la} \right) \right)^{-1}.
	\label{e:PB-phi-omega}
	\eeq
Moreover, this implies $\{ \tht, u \} =  (2 \om + m \la)/(k(2 \la \mathfrak{c} - 6 u))$, which notably differs from the original nilpotent Poisson bracket $\{ \tht, u \}_{\nu} = 0$ (\ref{e: PB-SL}).

\vspace{.25cm}

{\bf \fl Canonical action-angle variables on $\cal C$:} Since $\tht$ evolves linearly in time, it is a natural candidate for an angle variable. The corresponding canonically conjugate action variable $I$ must be a  function of $\omega, m$ and $\mathfrak{c}$ and is determined from (\ref{e:PB-phi-omega}) by the condition $\{ \tht , I(\omega) \} = I'(\omega) \{ \tht, \omega \} = 1$. We thus obtain, up to an additive constant,  the action variable 
	\beq
	I(\omega) = -k \om \left( 2 \mathfrak{c}  + \frac{3 m \omega}{\la} + \frac{2 \omega^2}{\la^2}  \right) = -k h.
	\label{e:circular-action-variable}
	\eeq
Thus we arrive at the remarkably simple conclusion that (aside from the Casimirs $\mathfrak{c}$ and $m$) $-k h$ and $\tht$ are action-angle variables on $\cal C$. Moreover, the canonical Poisson bracket $\{ \tht, -k h \} = 1$ agrees with that on the full phase space (see (\ref{e:PB-on-union-of-tori})). Our reason to work with $\om$ rather than $h$ as a coordinate is that the solutions (\ref{e:circular-level-set-solutions}) and the Hamiltonian (\ref{e:H-trigonometric}) have simple expressions in terms of $\om$. By solving the cubic (\ref{e:circular-action-variable}), $\om$ can be expressed in terms of $h$, which would allow us to write the Hamiltonian in terms of the action variable $- k h$.

\subsection{Union $\bar{\cal H}$ of horn toroidal level sets: Dynamics as gradient flow}
\label{s:horn-toroidal-level-sets}

Just as with the union of circular level sets $\cal C$, the union of horn toroidal level sets $\bar{\cal H}$ serves as the phase space for a self-contained dynamical system. However, unlike the sinusoidal periodic trajectories on $\cal C$, all solutions on $\bar{\cal H}$ are hyperbolic functions of time and are in fact homoclinic orbits joining the center of a horn torus to itself (see Fig.~\ref{f:theta-phi-dynamics-3D-horn-torus}). The centers themselves are static solutions. Horn tori arise only when the energy is equal to the critical value $E = E_{\rm sad}$ given in \S \ref{s:Hill-region-Morse-theory}. Thus, the horn tori are like the figure-8 shaped separatrices in the problem of a particle in a double well potential, separating two families of 2-tori. Interestingly, though the conserved quantities satisfy a relation on each horn torus, the four-fold wedge product $dh \wedge ds^2 \wedge dm \wedge d\mathfrak{c}$ vanishes only at its center. Finally, unlike on the circular submanifold, the flow on the horn-toroidal submanifold is {\it not} Hamiltonian, though we are able express it as a gradient flow.

The family of horn toroidal level sets is a two-dimensional submanifold ${\cal Q}_{\bar H}$ of the four-dimensional space of conserved quantities $\cal Q$. To see this, note that a horn torus arises when the cubic $\chi(u)$ of (\ref{e:cubic-equation-S3}) is positive between a simple zero and a double zero at the pole $u = s$ of the $S$-sphere. Thus, $\chi(u)$ must be of the form $\chi(u) = (u-u_1) (u-s)^2$ where $u_1 = \la m^2/2-s $ with $-s \leq u_1 \leq s$. These requirements imply $\chi(s) = \chi'(s) = 0$ and $\chi''(s) \geq 0$. Note that each non-trivial horn torus is a smooth two-dimensional surface except at its center which lies at the pole $u = s$. Trivial horn tori are those that have shrunk to the points at their centers and arise when $\chi''(s) = 0$. The conditions $\chi(s) = 0$ and  $\chi'(s) = 0$ lead to two relations among conserved quantities
	\beq
	h = -m s \quad \text{and} \quad \mathfrak{c} = \frac{m^2}{2} + \frac{s}{\la}, 
	\label{e:cons-qtys-Horn-tori-relations}
	\eeq
which together imply that $\D = 0$. The inequality $\chi''(s) \geq 0$ along with (\ref{e:cons-qtys-Horn-tori-relations}) restricts us to points above a parabola in the $m$-$s$ plane:
	\beq
	4s \geq \la m^2.
	\label{e:inequality-Horn-tori}
	\eeq 
The space ${\cal Q}_{\bar H}$ is given by  the set of such $(m,s)$ pairs. For each $(m,s) \in {\cal Q}_{\bar H}$ we get a horn torus $\bar H_{ms}$. The union of all horn tori is then given by $\bar {\cal H} = \cup_{4s \geq \la m^2} \bar H_{ms}$.  

\subsubsection{$\bar{\cal H}$ as a four-dimensional submanifold of $M^6_{S \text{-} L}$} 

Equations (\ref{e:cons-qtys-Horn-tori-relations}) and ({\ref{e:inequality-Horn-tori}}) when expressed in terms of $\vec S$ and $\vec L$ allow us to view the union of all horn tori $\bar {\cal H}$ as a four-dimensional submanifold of $M^6_{S \text{-} L}$:
	\beq
	S_1 L_1 + S_2 L_2 + ( S_3 - k s ) L_3 = 0, \quad \half (L_1^2 + L_2^2) + \frac{k S_3}{\la} = \frac{k^2 s}{\la} \quad \text{and} \quad L_3^2 \leq \frac{4 k^2 s}{\la}.
	\label{e:S-L-space-Horn-tori}
	\eeq
For any choice of $\vec S$, the first two conditions define a plane through the origin (normal to $(S_1, S_2, S_3 - s k)$) and a cylinder (of radius $r = \sqrt{(2 k/\la) (s k - S_3)}$ with axis along $L_3$) in the $L$-space. In general, this plane and cylinder intersect along an ellipse so that $\bar{\cal H}$ may be viewed as a kind of ellipse bundle over the $S$-space (subject to the inequality). The centers of the horn tori are the points where $S_{1,2} = L_{1,2} = 0$, $u = S_3/k = s$ and $|L_3/k| = |m| \leq \sqrt{4s/ \la}$ (see \S \ref{s:centers-of-horn-tori-punctured-horn-tori} below). Interestingly, it turns out that the inequality in (\ref{e:S-L-space-Horn-tori}) restricting the range of $L_3$ is automatically satisfied at all points of the base space other than when $u = s$ (which correspond to centers of horn tori). Indeed, let us find the range of values of $L_3$ allowed by the first two relations in (\ref{e:S-L-space-Horn-tori}) by parameterizing the elliptical fibre by the cylindrical coordinate $\tht$. Then $L_1 = r k \cos \tht$, $L_2 = r k \sin \tht$ and $L_3 = (2/ \la r) (S_1 \cos \tht + S_2 \sin \tht)$. The extremal values of $L_3$ on the ellipse occur at $\tht_{\rm ext} = \arctan S_2/S_1$ which implies that
	\beq
	|L_3|^2 \leq 
	\frac{2k}{\la} (sk + k \la) = \frac{4 k^2 s}{\la} - r^2.
	\eeq 
Thus the inequality in (\ref{e:S-L-space-Horn-tori}) is automatically satisfied away from the axis $r = 0$ which corresponds to the centers of horn tori.


\subsubsection{Centers of horn tori and punctured horn tori}
\label{s:centers-of-horn-tori-punctured-horn-tori}

It turns out that the centers of horn tori are static solutions and may therefore be regarded as forming the boundary of $\bar {\cal H}$. In particular, a trajectory on a horn torus $\bar H_{ms}$ can reach its center only when $t \to \pm \infty$. To find the space of centers $\cal O$ we note that they lie at the pole $u = s$ corresponding to $S_1 = S_2 = 0$ and $S_3 / k \geq 0$. The conditions (\ref{e:S-L-space-Horn-tori}) then become
	\beq
	(S_3 - k s) L_3 = 0, \quad \frac{L_1^2 + L_2^2}{2} + \frac{k S_3 }{\la} = \frac{k^2 s }{\la} \quad \text{and} \quad 4s \geq \la m^2 \quad \text{where} \quad s = \frac{S_3}{k}.
	\eeq 
The first condition is automatic, the second implies $L_{1,2} = 0$ while the inequality becomes $S_3 \geq (\la/4k) L_3^2$. Thus $\cal O$ is the two-dimensional subset of the static submanifold $\Sigma_2$ consisting of points on the $L_3$-$S_3$ plane, on or within the parabola $S_3 = (\la/4k) L_3^2$. The points on the parabola correspond to trivial horn tori. By eliminating their centers we obtain (non-trivial) punctured horn tori $H_{ms}$ which are smooth non-compact surfaces with the topology of infinite cylinders on which the dynamics is everywhere non static. We let ${\cal H} = \bar{\cal H} \setminus {\cal O} = \cup_{4s > \la m^2} H_{ms}$ denote the four-dimensional space consisting of the union of punctured horn tori. Thus $\cal H$ may be regarded as a cylinder bundle over the base ${\cal Q}_H = \{ (m, s) | 4s > \la m^2 \}$. Some possible coordinates on $\cal H$ are (a) $s, m, \tht, \phi$ (b) $s, m, u, \tht$ and (c) $S_{1,2,3}$ and either $L_1$ or $L_2$.

\subsubsection{Non vanishing four-fold wedge product on $\bar{\cal H}$} 

We have argued that the conserved quantities satisfy the relations  (\ref{e:cons-qtys-Horn-tori-relations}) on $\bar{\cal H}$. Despite this, {\it we show that the wedge product $\Om_4 = dh \wedge ds^2 \wedge dm \wedge d\mathfrak{c}$ does not vanish on $\bar {\cal H}$ except on its  boundary ${\cal O} = \bar{\cal H} \setminus {\cal H}$}. To see this, note that in addition to the condition $\D(\mathfrak{c}, m, h, s) = 0$ (due to the presence of the double zero at the pole $u = s$), all four partial derivatives of $\D$ may be shown to vanish on $\bar{\cal H}$ by virtue  (\ref{e:cons-qtys-Horn-tori-relations}). In other words, the relation $\D_\mathfrak{c} d\mathfrak{c} + \D_{m} dm + \D_{h} dh + \D_s ds = 0$ following from $\D = 0$ is vacuous on $\bar{\cal H}$ (if not, we could wedge it, say, with $ds^2 \wedge dm \wedge d\mathfrak{c}$ to show that $\Om_4 = 0$). On the other hand, we showed in \S 5.7 of \cite{G-V} that $\Om_4$ vanishes precisely on the closure of the circular submanifold $\bar{\cal C}= {\cal C} \sqcup {\cal C}_1 \sqcup {\cal C}_2 \sqcup (\Sigma_2 \cup \Sigma_3)$. Thus, to show that $\Om_4$ is non-vanishing on $\cal H$, it suffices to find the points common to $\bar{\cal H}$ and $\bar{\cal C}$. Now $\bar{\cal H} \cap {\cal C}$ is empty as $\chi$ has a double/triple zero at $u = s$ for points on $\bar{\cal H}$ and a double zero away from the poles for points on ${\cal C}$. In fact, we find that $\bar {\cal H} \cap \bar{\cal C}$ is contained in the static submanifold $\Sigma_2$ so that $\Om_4$ is nowhere zero on $\cal H$ and vanishes only on its boundary $\cal O$. To see that $\bar{\cal H}$ does not have any points in common with either ${\cal C}_1$ or  ${\cal C}_2$ we observe that the conditions $h = - ms$, $\mathfrak{c} = s/\la + m^2/2$ (\ref{e:cons-qtys-Horn-tori-relations}) and the relations ($S_1 = L_1 = 0$ and $\Xi_3$) or ($S_2 = L_2 = 0$ and $\Xi_2$) that go into the definitions of $\bar{\cal H}$ and ${\cal C}_1$ or ${\cal C}_2$ (see \S 5.6 of \cite{G-V}), together define a parabola in phase space
	\beq
	4 k S_3 = \la L_3^2 
	\quad \text{with} \quad k S_3 \geq 0 \quad 
	\text{and}  \quad L_{1,2} = S_{1,2} = 0.
	\eeq
This parabola is contained in $\Sigma_2$ but does not lie on ${\cal H}, {\cal C}_1$ or ${\cal C}_2$ as the inequalities $4s > \la m^2, |S_2| > 0$ and $|S_1| > 0$ appearing in the definitions of ${\cal H}, {\cal C}_1$ and ${\cal C}_2$ are saturated along it. Points on this parabola correspond to horn tori that have shrunk to the single point at their centers and correspond to cubics $\chi$ with a triple zero at $u=s$. Thus, this parabola lies along the common boundary of ${\cal H}, {\cal C}_1$ and ${\cal C}_2$. Combining these results we see that $\Om_4 \ne 0$ on $\cal H$, but vanishes identically on its boundary consisting of the space of centers $\cal O$.

\subsubsection{Equations of motion on the horn torus:} 

On the horn torus $H_{ms}$ the evolution equation for $u$ (\ref{e:EOM-u}) simplifies: 
	\beq
	\dot{u}^2 = 2 \la k^2 \chi(u) = \la^2 k^2 (s-u)^2 \left[\frac{2}{\la}(s + u)- m^2 \right].
	\eeq
We may interpret this equation as describing the zero energy trajectory of a non-relativistic particle of mass 2 with position $u(t)$ moving in a one-dimensional potential $V(u) = - 2 \la k^2 \chi(u)$. Since $V(u)$ is negative between the simple and double zeros at $u_1$ and $s$, the former is a turning point while the particle takes infinitely long to reach/emerge from $u=s$. Thus, the trajectory is like a solitary wave of depression. Choosing $u(0)$ to be its minimal value $u_1 = -s + \la m^2/2$, the trajectory of the particle is given by
	\beq
	u(t) = u_1 + (s-u_1) \tanh^{2} \left(\frac{t}{2\tau}\right)
	\quad \text{where} \quad \tau = \ov{\sqrt{\la k^2 (4s - \la m^2)}}.
	\label{e:particle-trajectory-horn-torus}
	\eeq
Notice that as $t \to \pm \infty$, $u(t) \to s$ and the solution approaches the center of the horn torus. Interestingly, the vector field $\dot u = \sqrt{-V(u)}$ is not smooth at $u = u_1$, which is a square root branch point. Thus, there is {\it another} solution $u(t) \equiv u_1$ with the same initial condition (IC) $u(0) = u_1$, which however is consistent with the $L$-$S$ equations of motion (\ref{e: EOM-LS}) only when $s = 0$. Note that (\ref{e:particle-trajectory-horn-torus}) can be obtained  as a limit of the $\wp$-function solution given in \S \ref{s:Reduction-2D}. On a horn torus, one of the half periods of the $\wp$-function is imaginary while the other diverges leading to the aperiodic solution (\ref{e:particle-trajectory-horn-torus}). 

\begin{figure}[h]
	\centering
		\begin{subfigure}[t]{5cm}
		\centering
		\includegraphics[width=5cm]{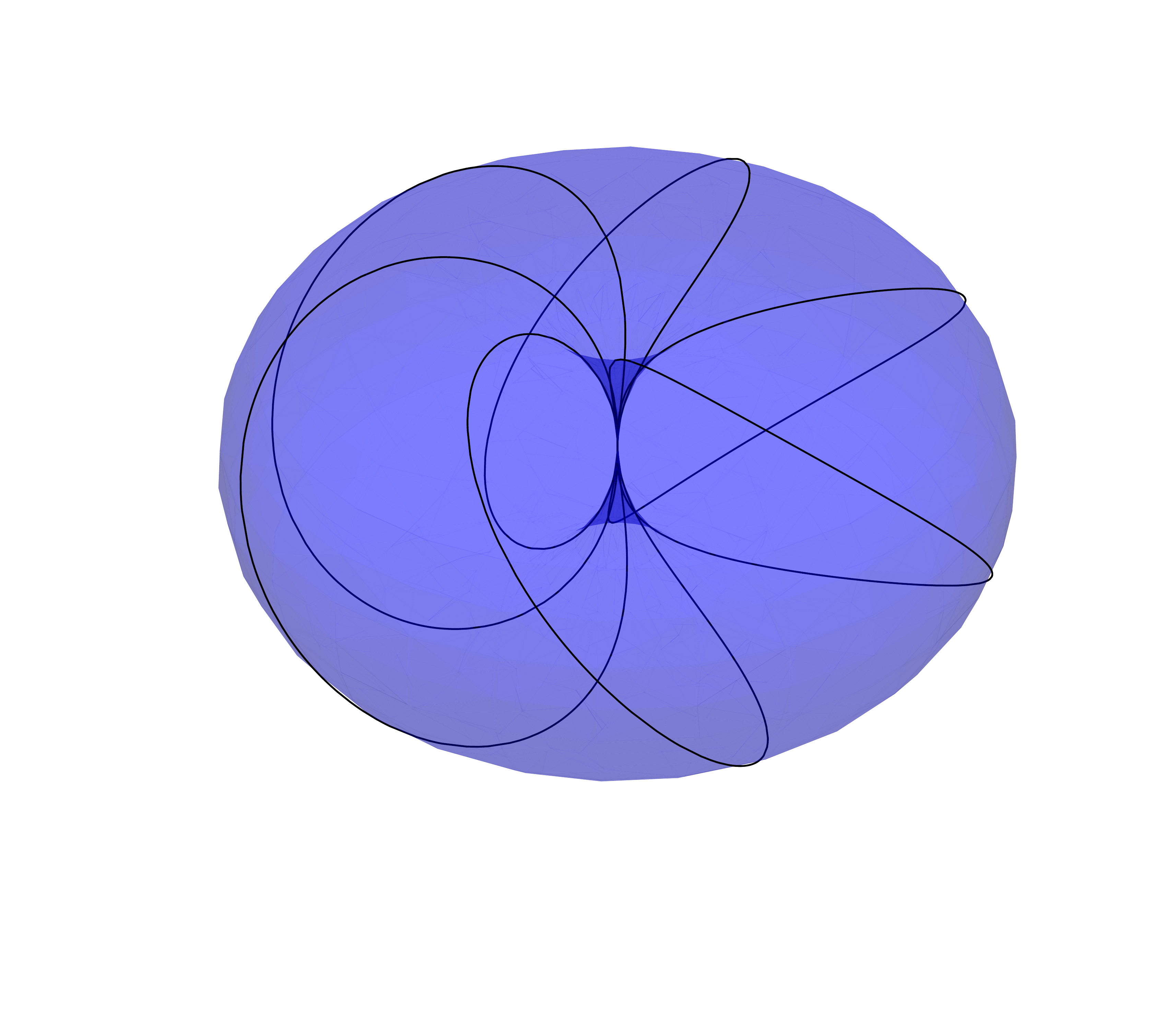}
		\caption{}
		\label{f:horn-torus-trajectories}
		\end{subfigure}
		\qquad \quad
		\begin{subfigure}[t]{3cm}
		\centering
		\includegraphics[width= 3cm]{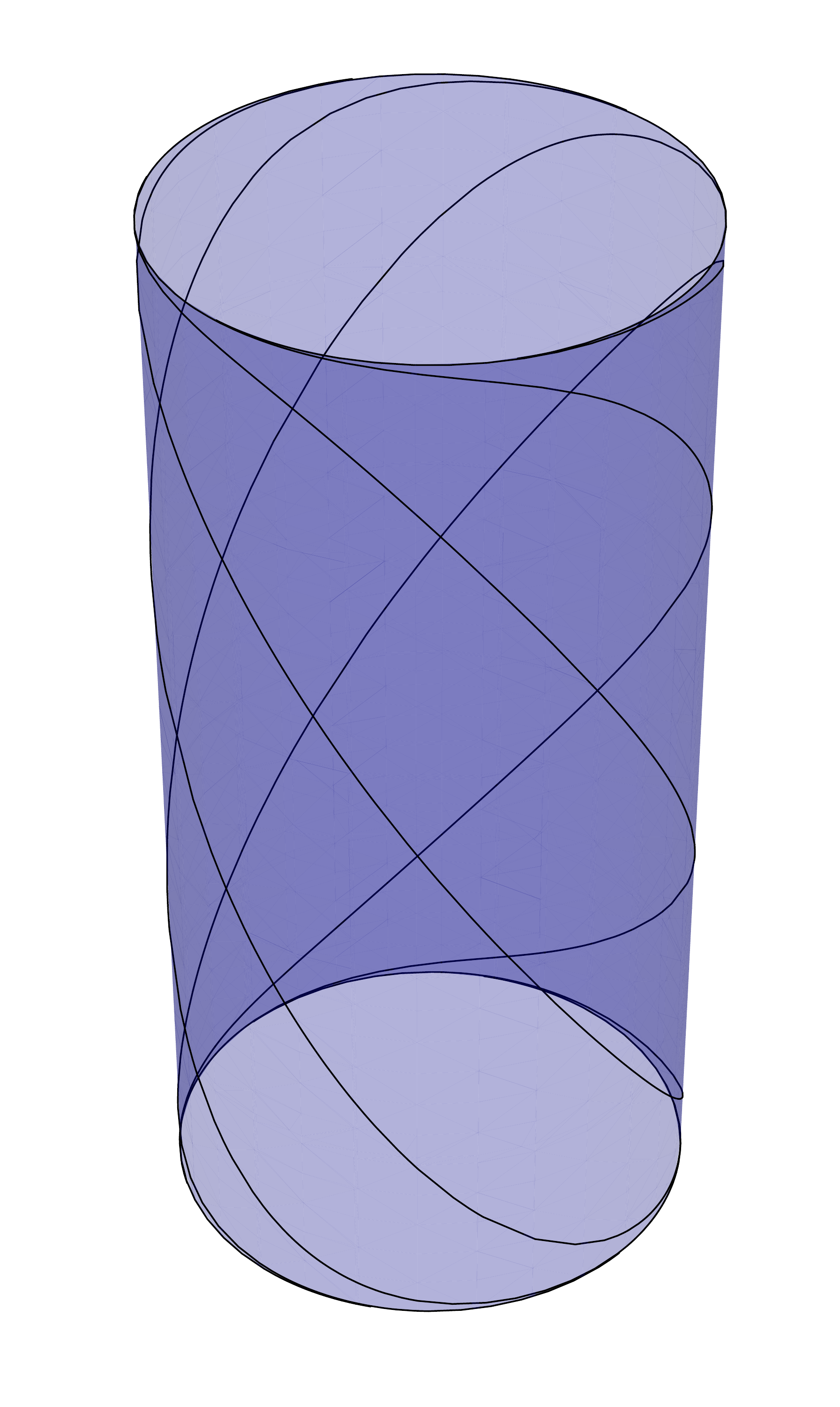}
		\caption{}
		\label{f:theta-phi-integration-cylinder-horn-torus}
		\end{subfigure}
	\caption{\footnotesize Six trajectories on a punctured horn torus (with $s = 1, m = -1$ and $\la = k = 1$) displayed in two embeddings [(a) Eq. (\ref{e:horn-torus-embedding}) and (b) Eq. (\ref{e:horntorus-cylinder-embedding}) with $R = 1.5$] passing through the points $\tht_e(0)= 0$ and $\phi_e(0) = 0, \pi/3, 2\pi/3, \pi, 4\pi/3, 5\pi/3$ extended indefinitely forward and backward in time. Trajectories emerge from the center (at $t = -\infty$) and approach the attractor at the center as $t \to \infty$ showing that the phase space volume cannot be preserved. In (b), the top and bottom rims of the cylinder correspond to the center of the horn torus.}
	\label{f:theta-phi-dynamics-3D-horn-torus}
\end{figure}

To describe the trajectories on a horn torus $H_{m s}$ we use the coordinates $\tht = \arctan(L_2/L_1)$ and $\phi = \arctan(S_2/S_1)$ in terms of which the equations of motion (\ref{e:theta-phi-dynamics}) simplify to
	\beq
	\dot \tht = \frac{k m \la}{2} \quad \text{and} \quad \dot \phi = \frac{k m \la s}{s+u} = \frac{2 k s \cos^2(\tht - \phi)}{m}. 
	\eeq
Notice that $\tht$ is monotonic in time: increasing/decreasing according as $\sign{km} = \pm 1$. It is convenient to pick ICs on the curve $u = u_1$ resulting in the solution
	\beq
	\tht(t) = \tht(0) + \frac{k m \la t}{2} \quad \text{and} \quad 
	\phi(t) = \phi(0) + \frac{k m \la t}{2} + 
	\arctan\left(\frac{ \tanh\left(\frac{t}{2\tau}\right)}{k \tau m \la}\right).
	\label{e:theta-phi-dynamics-horn-torus}
	\eeq
Though $\tht$ and $\phi$ are both ill-defined at the center of the horn torus ($L_{1,2} = S_{1,2} = 0$), we notice from (\ref{e:relation-theta-phi-u}) that the difference $\tht-\phi$ is well defined at the center: 
	\beq
	\lim_{t \to \pm \infty} (\tht(t) - \phi(t)) = \arccos \sqrt{\frac{\la m^2}{4s}} = \lim_{u \to s} (\tht - \phi).
	\eeq	
Since $\tht$ is ill-defined at the center $u=s$, it is convenient to switch to the `embedding' variables:
	\beq
	\tht_e = \frac{\pi (\phi - \tht)}{\sign{m k} \arctan \left(1/ m \la k \tau \right)}  \quad \text{and} \quad \phi_e = \phi.
	\eeq
The advantage of $\tht_e$ is that it approaches $\pm \pi/\sign{mk}$ as $t \to \pm \infty$ on any trajectory on $H_{ms}$. We may visualize the dynamics via the following embedding of the horn torus in Euclidean 3-space:
	\beq
	x = R(1+\cos \tht_e) \cos \phi_e, \quad y = R(1+\cos \tht_e) \sin \phi_e \quad \text{and} \quad z = R \sin \tht_e.
	\label{e:horn-torus-embedding}
	\eeq	
Here $R$ is the major (as well as the minor) radius of the horn torus (see Fig.~\ref{f:horn-torus-trajectories}). Alternatively, we may realize the punctured horn torus as a cylinder in three-dimensional space via the embedding
	\beq
	x = R \cos \phi_e, \quad y = R \sin \phi_e \quad \text{and} \quad z = \tht_e.
	\label{e:horntorus-cylinder-embedding}
	\eeq 
The center of the horn torus lies at $\tht_e = \pm \pi \: (\text{mod} \: 2\pi)$ with $\phi_e$ arbitrary (see Fig.~\ref{f:theta-phi-integration-cylinder-horn-torus}). As $t \to \pm \infty$ all trajectories spiral into the center of the horn torus as shown in Fig. \ref{f:theta-phi-dynamics-3D-horn-torus}. Thus, every trajectory is homoclinic, beginning and ending at the center of the horn torus.

As noted in \S \ref{s:Hill-region-Morse-theory}, horn tori arise only at the saddle points of the Hamiltonian $H = k^2 E_{\rm sad}$. Thus, they are analogs of the figure-8 shaped separatrix at energy $g a^4$ familiar from particle motion in the one-dimensional potential $V(x) = g(x^2 - a^2)^2$. For fixed $\mathfrak{c}, m$ with $2 \mathfrak{c} - m^2 > 0$, and $E = E_{\rm sad}$, $h$ can take a range of values from $h_{\rm min}$ to $h_{\rm max}$. There is a critical value $h_{\rm sad}$  in this range at which the common level set is a horn torus. It is flanked by 2-tori on either side. Thus, horn tori separate two families of toroidal level sets with the real half-period $\om_R$ of the $\wp$-function diverging as $h \to h_{\rm sad}^{\pm}$. 

\subsubsection{Flow on $\cal H$ is not Hamiltonian}

 The equations of motion on $\cal H$		
	\beq
	\dot s = \dot m = 0, \quad 
	\dot \tht = \half k m \la \quad \text{and} \quad 
	\dot \phi = \frac{k m \la s}{s + u} = \frac{2 k s \cos^2(\tht - \phi)}{m},
	\label{e:EOM-horn-tori}
	\eeq
 do not follow from any Hamiltonian and Poisson brackets on $\cal H$. This is because time-evolution does not satisfy the Liouville property of preserving phase volume: every initial condition is attracted to the center of a horn torus. Said differently, the flow can map a subset $I_0$ of $\cal H$ into a proper subset $I_t \subsetneq I_0$. To show this, it suffices to consider the dynamics on each $H_{ms}$ separately since the dynamics preserves individual punctured horn tori. Thus, consider the `upper cylinder' subset of $H_{ms}$: $I_0 = \{ (\phi_e, \tht_e)| \: \tht_e \geq \tht_0 \; \text{for some} \: -\pi < \tht_0 < \pi \}$. Then
 	\beq
	I_t = \left\{(\phi_e, \tht_e)| \tht_e > \tht_0 - \frac{\pi(\tht(t)-\phi(t))}{\sign{km} \arctan(1/k \tau m \la)} \right\}
	\eeq
is its image under evolution to time $t$. Since $\tht_e$ is monotonic in time, we observe that for $km > 0$, $I_t$ form a 1-parameter family of subsets with decreasing volume (relative to any reasonable volume measure on $H_{ms}$) while $\text{vol}(I_{t})$ grows if $km < 0$. Thus, the Liouville theorem would be violated if the dynamics on $H_{ms}$ or $\cal H$ were Hamiltonian. 

Interestingly, time evolution on $\cal H$ may be realized as a gradient flow. As before, we focus on the dynamics on each $H_{ms}$ separately. Since $W = -\sign{km} \tht$ is monotonically decreasing in time (\ref{e:EOM-horn-tori}), we choose it as the potential function for the gradient flow
	\beq
	\dot \xi^i = (\dot{\phi}, \dot \tht ) = V^i(\xi) = -g^{ij} \dd{W}{\xi^j} \quad
	\text{where} \quad  
	V^\phi = \frac{2 k s \cos^2(\tht - \phi)}{m} \quad \text{and} \quad
	V^\tht = \frac{k m \la}{2}.
	\eeq
	The inverse-metric on $H_{ms}$ that leads to this gradient flow must be of the form
	\beq
	g^{ij} = \sign{k m}\colvec{2}{\Upsilon & \dot{\phi}}{\dot{\phi} & \dot{\tht}}. 
	\eeq
Here $\Upsilon$ is an arbitrary function on $H_{ms}$ which we may choose so that the metric is, for simplicity, Riemannian (positive definite). This is ensured if
	\beq
	\det{g^{-1}}  > 0  \quad \Leftrightarrow \quad \Upsilon \dot{\tht} > \dot{\phi}^2 \quad \text{and} \quad \tr{g^{-1}} > 0 \quad \Leftrightarrow \quad \sign{k m} (\Upsilon + \dot{\tht} ) > 0.
	\eeq
The second condition is implied by the first, so a simple choice that ensures a Riemannian metric is $\Upsilon = (\dot{\phi}^2 / \dot{\tht}) + \sign{km}\: \epsilon$, for any $\epsilon > 0$. It might come as a surprise that this gradient flow admits homoclinic orbits beginning and ending at the center.
Such orbits are typically forbidden in gradient flows. Our horn tori evade this `no-go theorem' since the potential $W \propto \tht$ is not defined at the centers of horn tori.

\subsection{Dynamics on the union $\cal T$ of toroidal level sets}
\label{s:toroidal-level-sets}

For generic values of $\mathfrak{c}, m, s$ and $h$, i.e., for which the discriminant $\D \ne 0$ (\ref{e:discriminant}), the common level sets are 2-tori as shown in \S \ref{s:canonical-vector-fields-topology} and \S \ref{s:common-level-set-conserved-qtys}. The union  $\cal T$ of these 2-tori may be viewed as the state space of a self-contained dynamical system. Here, we express $\cal T$ as a torus bundle over a space ${\cal Q}_{\cal T}$ of conserved quantities, and find a convenient set of local coordinates on it along with their Poisson brackets implied by (\ref{e: PB-SL}). We use this Poisson structure and the time evolution of $u$ in terms of the $\wp$ function (\ref{e:u-wp-function}) to find a family of action-angle variables on $\cal T$. Finally, we  show that these action-angle variables degenerate to those on the union $\cal C$ of circular level sets when the tori degenerate to circles.

\subsubsection{Union of toroidal level sets}

Let us denote by ${\cal Q}_{\cal T}$, the subset $\D(\mathfrak{c}, m, s, h) \neq 0$ of the space $\cal Q$ of conserved quantities for which the common level sets are 2-tori. On ${\cal Q}_{\cal T}$ the cubic $\chi(u)$ (\ref{e:cubic-equation-S3}) is positive between two adjacent simple zeros $u_{\rm min}$ and $ u_{\rm max}$ and the common level set $M^{s h}_{\mathfrak{c} m}$ is a torus. Thus, on ${\cal Q}_{\cal T}$ the cubic takes the form $\chi(u) = (u - u_{\rm min})(u_{\rm max} -u)(u_3 - u)$ with $-s \leq u_{\rm min} < u_{\rm max} \leq s$ and $u_{\rm max} < u_3$. In this case, when $\chi(u)$ is written in Weierstrass normal form using $u = a v + b$, the   invariants $g_2$ and $g_3$ are real and the discriminant of the cubic is non-zero. It follows that the half periods $\om_R$ and $\om_I$ of \S \ref{s:Reduction-2D} are respectively real and purely imaginary. We designate the union of these tori ${\cal T} \subset M^6_{S \text{-} L}$ and the corresponding union for fixed $\mathfrak{c}$ and $m$, ${\cal T}^4_{\mathfrak{c} m}$. Here, ${\cal T}$ may be visualised as a torus bundle over ${\cal Q}_{\cal T}$. While $\tht$ and $\phi$ furnish global coordinates on the torus $M^{s h}_{\mathfrak{c} m}$, it is more convenient, when formulating the dynamics, to work with the local coordinates $(u, \tht)$ where $\cos (\tht - \phi) = (h + m u)/r \rho$. An advantage of $u$ is that unlike $\phi$, it commutes with $h$. However, since the cosine is a 2:1 function on $[0,2\pi]$, we need two patches $U_\pm$ with local coordinates $(u_\pm, \tht)$ to cover the torus with $u_{\rm min} \leq u_\pm \leq u_{\rm max}$ and $0 \leq \tht \leq 2 \pi$. In the $U_\pm$ patches, the formula for $\phi$ is
	\beq
	\phi = \tht \pm \arccos \left( \frac{h + m u}{r \rho}\right)_{[0, \pi]},
	\eeq
where the $\arccos$ function is defined to take values between $0$ and $\pi$. Whenever $u$ reaches either $u_{\rm min}$ or $u_{\rm max}$, the trajectory crosses over from one patch to the other.

\begin{figure}[h]
	\centering
		\begin{subfigure}[b]{5cm}
		\includegraphics[width=5cm]{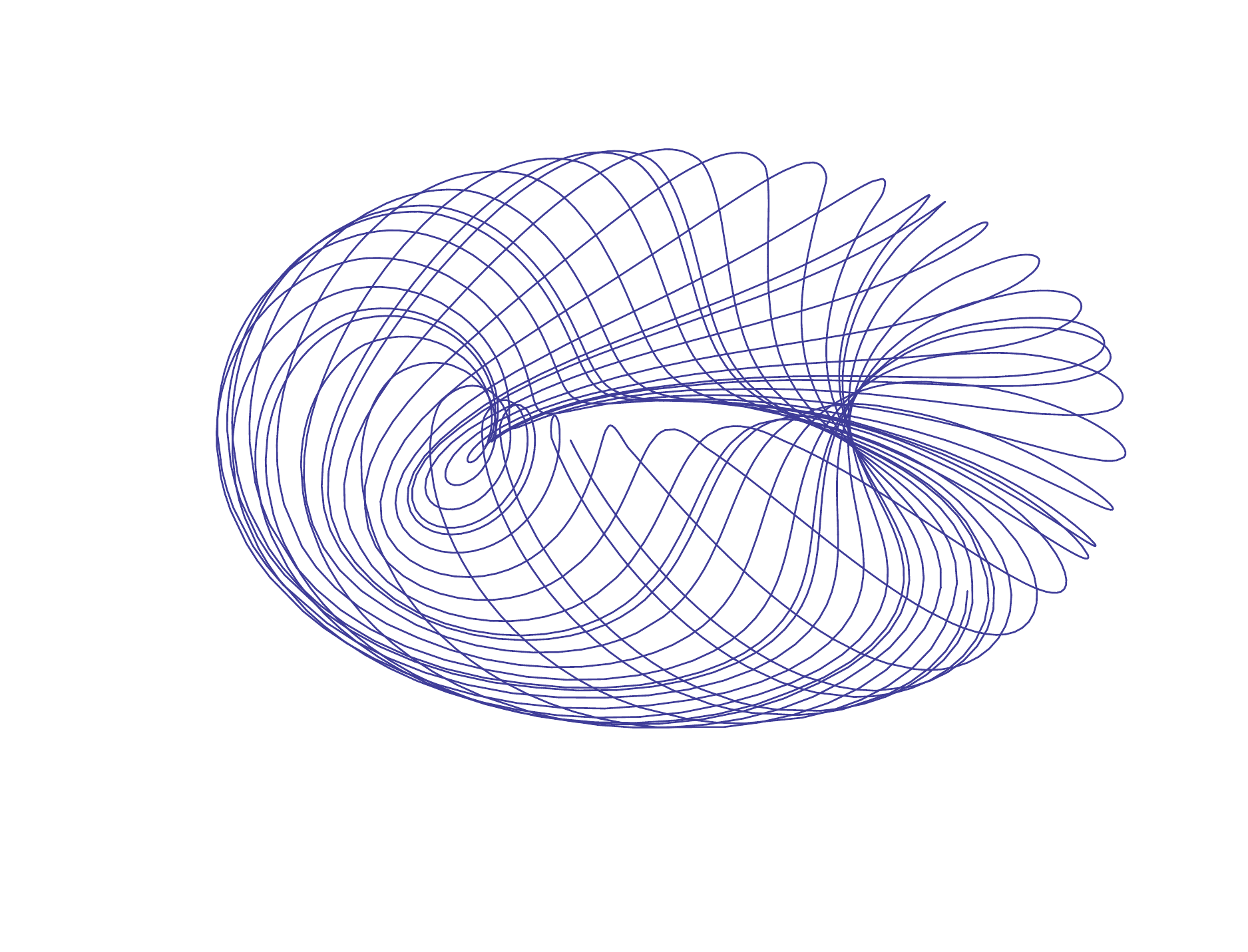}
		\end{subfigure}
	\caption{\footnotesize Trajectory on an invariant torus for the parameters $k = \la = 1, \mathfrak{c} = 3, h= 1, m = -1, s=1$ and $R = 2$ for $0 < t < 75 \om_R$ ($\om_R \approx 1.41$ is the real half-period of $u$ (\ref{e:u-wp-function})) displayed via the embedding $x = (R + \varrho \cos\tht_e)\cos \phi_e$, $y = (R + \varrho \cos \tht_e) \sin \phi_e$ and $z = \varrho \sin \tht_e$.  The poloidal and toroidal angles are $\tht_e =  \arcsin \left( (u - \bar{u})/\varrho  \right)$ and $\phi_e = \phi$ with $\bar{u} = (u_{\rm min} + u_{\rm max})/2$ and $\varrho = (u_{\rm max} - u_{\rm min})/2$. Unlike the angle variables $\tht^1$ and $\tht^2$ (\ref{e:action-angle-variables-torus}),  which are periodic on account of their linearity in time, neither $\tht_e$ nor $\phi_e$ is periodic.}
	\label{f:torus-plot-theta-phi}
\end{figure}


\subsubsection{Poisson structure on $\cal T$}

On ${\cal T}$, we use the local coordinates $\mathfrak{c}, m, s, h, \tht$ and $u$. The Poisson structure following from the nilpotent Poisson brackets (\ref{e: PB-SL}) is degenerate with the Casimirs $\mathfrak{c}$ and $m$ generating the center. The Poisson brackets among the remaining coordinates (on ${\cal T}^4_{\mathfrak{c} m}$) are:
	\beqs
	\{ s, h \} &=& \{ h, u \} = \{ \tht, u \} = 0 , \quad \{ h, \tht \} = \ov k, \quad  
	\{ s, \tht \} = \frac{h + m u}{k s r^2} = \frac{\rho}{k s r} \cos(\tht - \phi) = -\frac{\dot{\tht}}{k^2 s}, \cr
	\{ s, u_{\mp} \} &=& \mp \frac{\la}{k s} \sqrt{r^2 \rho^2 - (h + m u)^2} = \mp \frac{ \sqrt{2 \la k^2 \chi(u)}}{k^2 s} = -\frac{r \rho \la}{k s} \sin(\tht - \phi) = -\frac{\dot{u}}{k^2 s}.
	\label{e:PB-on-union-of-tori}
	\eeqs	
All the Poisson brackets other than $\{ s, u \}$ have a common expression on both patches $U_{\pm}$. Here $r^2 = 2 \mathfrak{c} - m^2 - 2u/\la$ and $\rho^2 = s^2 - u^2$. 

\subsubsection{Action-angle variables on $\cal T$}

We seek angle-action variables $(\tht^1,\tht^2, I_1,I_2)$ on ${\cal T}^4_{\mathfrak{c} m}$ satisfying canonical Poisson brackets
	\beq
	\{ \tht^i, \tht^j \} = \{ I_i , I_j \} = 0 \quad \text{and} \quad \{ \tht^i, I_j \} = \delta^i_j.
	\label{e:canonical-act-ang-PB}
	\eeq
The action variables $I_1$ and $I_2$ must be conserved and therefore functions of $s$ and $h$ alone, while the angles $\tht^1$ and $\tht^2$ must evolve linearly in time: $\dot \tht^j = \Om_j(s,h)$. Here we suppress the parametric dependence of $\tht^i$ and $I_j$ on the Casimirs $\mathfrak{c}$ and $m$ which specify the symplectic leaf. In what follows, we use the $\wp$-function solution (\ref{e:u-wp-function}) along with the requirement of canonical Poisson brackets to find a family of action-angle variables. Despite some long expressions in the intermediate steps, the final formulae  (\ref{e:action-angle-variables-torus}) for $(\tht^i, I_j)$ are relatively compact. Though we work here with the nilpotent Poisson structure (\ref{e:PB-on-union-of-tori}), it should be possible to generalize the resulting action-angle variables to the other members of the Poisson pencil (\ref{e:Poisson-pencil}).  

\vspace{.25cm}

{\fl \bf Determination of $\tht^1$ and $I_1$:} The evolution of $u$ (\ref{e:u-wp-function}) gives us one candidate for an angle variable evolving linearly in time 
	\beq
	\tht^1 =  k\left( \wp^{-1}\left(\frac{k^2 \la}{2} \left(u - \frac{\mathfrak{c} \la}{3} \right); g_2, g_3 \right) - \alpha(\mathfrak{c}, m, s, h) \right)  = k( t + t_0).
	\label{e:angle-variable-theta-1}
	\eeq
The factor of $k$ is chosen to make $\tht^1$ dimensionless. Here, $g_2$ and $g_3$ (\ref{e:Weierstrass-invariants}) are functions of the conserved quantities. From the definition of $\tht^1$, it follows that the frequency $\Omega_1 = k$. Choosing $\al$ to be the imaginary half-period $\om_I$ of the $\wp$-function in (\ref{e:u-wp-function}) ensures that $\tht^1$ is real. An action variable conjugate to $\tht^1$ is 
	\beq
	I_1(s, h) = \frac{k s^2}{2}  + f(h),
	\label{e:I1-action-avariable}
	\eeq
where $f'(h) \neq 0$ is an arbitrary function of $h$ (and possibly $\mathfrak{c}$ and $m$) to be fixed later. Upto the function $f$, $I_1$ is proportional to the Hamiltonian (\ref{e:Hamiltonian-s}). Eq. (\ref{e:I1-action-avariable}) is obtained by requiring
	\beq
	\{\tht^1, I_1 \} = \dd{\tht^1}{u} \dd{I_1}{s} \{ u, s \} = k\dd{(\wp^{-1}(v) - \al)}{v} \dd{v}{u}\dd{I_1}{s} \{ u, s \} = \frac{k}{\dot{u}} \dd{I_1}{s} \frac{\dot{u}}{k^2 s} = 1.
	\eeq
Here, $v = (u-b)/a$ (see \S \ref{s:Reduction-2D}) and we used the relation
	\beq
	\dd{\wp^{-1}(v; g2, g3)}{v} = \frac{1}{\dot{v}} = \frac{a}{\dot{u}}.
	\label{e:derivative-wp-inverse}
	\eeq 
For future reference we also note that as a consequence, $\pdr \tht^1/ \pdr u = k/ \dot{u}$. This derivative diverges at $u_{\rm min}$ and $u_{\rm max}$, which are the roots of $\chi$.

\vspace{.25cm}

{\fl \bf Determination of $\tht^2$ and $I_2$:} To identify the remaining action-angle variables $I_2(s,h)$ and $\tht^2(u,\tht,s,h)$ we first consider the constraints coming from the requirement that their Poisson brackets be canonical. While $\{ I_1, I_2 \} = 0$ is automatic, $\{ \tht^1 , I_2 \} = 0$ implies that $I_2(s,h)$ must be independent of $s$:
	\beq
	0 = \{ \tht^1 , I_2 \} = \dd{I_2}{s} \{ \tht^1, s \} + \dd{I_2}{h} \cancel{\{ \tht^1, h \}} \quad \imply \quad \dd{I_2}{s} = 0. 
	\eeq
The remaining Poisson brackets help to constrain $\tht^2$. For instance, $\{ \tht^2, I_2(h) \} = 1$ forces $\tht^2$ to be a linear function of $\tht$:
	\beq
	\{ \tht^2, I_2(h) \} = \dd{\tht^2}{\tht} I_2'(h) \{ \tht, h \} = -\dd{\tht^2}{\tht} \frac{I_2'(h)}{k} = 1  \quad \imply \quad \tht^2 =  -\frac{k}{I_2'(h)} \tht + g(u,s,h).
	\label{e:tht2-I2-PB}
	\eeq
Here $g$ is an arbitrary function which we will now try to determine. Next, $\{ \tht^2, I_1 \} = 0$ implies that $\tht^2$ evolves linearly in time: 
	\beqs	
	\{ \tht^2, I_1 \} = \dd{\tht^2}{u} \{ u, I_1 \} + \dd{\tht^2}{\tht} \{ \tht, I_1 \} = 0 \quad &\imply& \quad   \dot{\tht^2} =   \dd{\tht^2}{u} \dot{u} + \dd{\tht^2}{\tht} \dot{\tht}  =  f'(h) \dd{\tht^2}{\tht} \equiv \Omega_2 \cr
	&\imply& \quad \tht^2 = \frac{\Omega_2}{f'(h)} \tht + g(u,s,h).
	\label{e:linear-time-theta-2}
	\eeqs	
Comparing (\ref{e:tht2-I2-PB}) and (\ref{e:linear-time-theta-2}), it follows that $ \Omega_2 = -k f'(h)/I_2'(h)$ is independent of $s$. We may use (\ref{e:linear-time-theta-2}) to reduce the determination of the dependence of $\tht^2$ on $u$ to quadratures: 
	\beq
	\dot{\tht^2} = \Omega_2 = \frac{\Omega_2}{f'(h)} \dot{\tht} + \dd{g(u,s,h)}{u} \dot{u}.
	\label{e:time-evolution-tht-2}
	\eeq
Using (\ref{e:theta-phi-dynamics}) and (\ref{e:EOM-u}) we get
	\beq
	\dd{\tht^2}{u} = \dd{g}{u} 
	= \pm\Om_2  \frac{1+ \frac{k}{f'(h)}\left(\frac{h + m u}{2\mathfrak{c} - m^2 - {2 u}/{\la}}\right)}{\sqrt{2 \la k^2 \chi(u)}}.
	\label{e:u-dependence-tht-2}
	\eeq
Integrating,
\small
	\beq
	\frac{g(u, s, h)}{\Om_2} = \frac{ \pm 1}{\sqrt{2 \la k^2}} \left[ \left(1 - \frac{k m \la}{2 f'(h)} \right) \int_{u_{\rm min}}^{u} \frac{du'}{\sqrt{\chi(u')}} - \frac{k m \la}{2 f'(h)}\left( \frac{h}{m} + u_0 \right) \int_{u_{\rm min}}^u \frac{du'}{\left( u' - u_0 \right) \sqrt{\chi(u')}} \right]  + \tl{g}(s, h). \qquad
	\label{e:u-dependence-tht-2-integral}
	\eeq
\normalsize
where $u_0 = \mathfrak{c}/ \la - m^2 \la/ 2$. Recognizing these as incomplete elliptic integrals of the first and third kinds ($F$ and $\Pi$), we get (see \S 3.131, Eq. (3) and \S 3.137, Eq. (3) of \cite{G-R})
\small
	\beq
	\frac{g}{\Om_2} = \pm \sqrt{\frac{2}{\la k^2}}\frac{k m \la}{2 f'(h)} \left[ \left( \frac{2 f'(h)}{k m \la} - 1 \right) \frac{F(\g , q)}{\sqrt{u_3 - u_{\rm max}}} + \left( \frac{h}{m} + u_0 \right) \frac{\Pi \left( \gamma, \frac{u_{\rm max} - u_{\rm min}}{u_0 - u_{\rm min}}, q \right)}{(u_0 - u_{\rm min}) \sqrt{u _3 - u_{\rm min}} }  \right] + \tl{g}(s, h). 
	\label{e:angle-variable-elliptic-integral}
	\eeq
\normalsize
Here, $\tl{g}(s,h)$ is an integration constant, $u \in [u_{\rm min},u_{\rm max}]$ where $-s \leq u_{\rm min} < u_{\rm max}  < u_3$ (which are functions of $\mathfrak{c}, m, s$ and $h$) are the roots of the cubic $\chi(u)$. Moreover, the amplitude and elliptic modulus are
	\beq
	\gamma = \arcsin \sqrt{\frac{u - u_{\rm min}}{u_{\rm max} - u_{\rm min}}} \quad \text{and} \quad 
	q = \sqrt{\frac{u_{\rm max} - u_{\rm min}}{u_3 - u_{\rm min}}}.
	\eeq
To find the $s$ dependence of $\tht^2$, we notice that the last Poisson bracket $\{ \tht^1, \tht^2 \} = 0$ gives the following relation among derivatives of $\tht^2$:
	\beqs
	\{ \tht^1 , \tht^2 \} &=& \dd{\tht^2}{u} \{ \tht^1 , u \} + \dd{\tht^2}{\tht} \{ \tht^1 , \tht \} + \dd{\tht^2}{s} \{ \tht^1 , s \}  + \dd{\tht^2}{h} \cancel{\{ \tht^1 , h \}} = 0 \cr
	\imply  && \dd{\tht^2}{\tht} \dd{\tht^1}{s} \{ s, \tht \}  + \dd{\tht^2}{\tht} \dd{\tht^1}{h} \{ h , \tht \}  + \left( \dd{\tht^2}{u} \dd{\tht^1}{s} - \dd{\tht^2}{s} \dd{\tht^1}{u} \right) \{ s, u \} = 0.  
	\label{e:-tht1-tht2-PB-torus}
	\eeqs
Using the known formulae for the partial derivatives (\ref{e:angle-variable-theta-1}, \ref{e:derivative-wp-inverse}, \ref{e:linear-time-theta-2}, \ref{e:time-evolution-tht-2})
	\beq
	\frac{\partial \tht^2}{\partial \tht} = \frac{\Om_2}{f'(h)}, \quad 
	\dd{\tht^2}{u} = \frac{\Om_2}{\dot u}\left( 1 - \frac{\dot \tht}{f'(h)}\right), \quad
	\dd{\tht^1}{u} = \frac{k}{\dot u} \quad \text{and} \quad
	\pdr_{s,h} \tht^1 = k \pdr_{s, h} (\wp^{-1} - \om_I), 
	\eeq
we find the $s$ dependence of $\tht^2$ from (\ref{e:-tht1-tht2-PB-torus}):
	\beq
	\dd{\tht^2}{s} =\Om_2(h) \left[ \pdr_{s} (\wp^{-1} - \om_I) - \frac{k s}{f'(h)} \pdr_{h} (\wp^{-1} - \om_I) \right].
	\label{e:s-dependence-theta-2}
	\eeq
In effect, we have two expressions ((\ref{e:angle-variable-elliptic-integral}) and (\ref{e:s-dependence-theta-2})) for $\pdr_s\tht^2$. We exploit them to reduce the determination of the $s$ dependence of $\tht^2$ to quadrature. Comparing $\pdr_s$(\ref{e:angle-variable-elliptic-integral}) with (\ref{e:s-dependence-theta-2}) gives 
\footnotesize
	\beqs
	\pdr_s {\tl g} &=& \frac{\pdr}{\pdr s} \left[ \wp^{-1} - \om_I \mp \sqrt{\frac{2}{\la k^2}}\frac{k m \la}{2 f'(h)} \left\{ \left( \frac{2 f'(h)}{k m \la} - 1 \right) \frac{F(\g , q)}{\sqrt{u_3 - u_{\rm max}}} + \left( \frac{h}{m} + u_0 \right) \frac{\Pi \left( \gamma, \frac{u_{\rm max} - u_{\rm min}}{u_0 - u_{\rm min}}, q \right)}{(u_0 - u_{\rm min}) \sqrt{u _3 - u_{\rm min}} }  \right\} \right] \cr
	&& - \frac{k s}{f'(h)} \pdr_h(\wp^{-1} - \om_I).
	\eeqs
\normalsize
Thus
\footnotesize
	\beqs
	\tl{g}(s,h) &=& \wp^{-1} - \om_I \mp \sqrt{\frac{2}{\la k^2}}\frac{k m \la}{2 f'(h)} \left[ \left( \frac{2 f'(h)}{k m \la} - 1 \right) \frac{F(\g , q)}{\sqrt{u_3 - u_{\rm max}}} + \left( \frac{h}{m} + u_0 \right) \frac{\Pi \left( \gamma, \frac{u_{\rm max} - u_{\rm min}}{u_0 - u_{\rm min}}, q \right)}{(u_0 - u_{\rm min}) \sqrt{u _3 - u_{\rm min}} }  \right] \cr
	&& - \int^s_{\infty} \frac{k s'}{f'(h)} \pdr_h(\wp^{-1} - \om_I) \: ds' + \eta(h).
	\label{e:s-h-dependence-of -tht-2}
	\eeqs
\normalsize
Here $\eta(h)$ is an arbitrary `constant' of integration. Now, using (\ref{e:s-h-dependence-of -tht-2}) in (\ref{e:angle-variable-elliptic-integral}) results in some pleasant cancellations leading to a relatively simple formula for $g$:
	\beq
	\frac{g(u,s,h)}{\Om_2} = \wp^{-1} - \om_I - \frac{k}{f'(h)} \int_{\infty}^s  s' \:\pdr_h (\wp^{-1} - \om_I) \: ds' + \eta(h).
	\label{e:g-u-s-dependence-of-tht-2}
	\eeq
This determines the angle variable $\tht^2(\tht, u, s, h) = \Om_2 \tht/ f'(h) + g(u, s, h)$. It is noteworthy that $ \wp^{-1} - \om_I$ is simply $\tht^1/k$. The integral over $s'$ is from $\infty$ since, for sufficiently large $s$, $\D$ (\ref{e:discriminant}) is always positive so that $M_{\mathfrak{c} m}^{s h}$ is a torus. However, we must take $s > s_{\rm min}$, which is the value at which $\D$ vanishes and the torus $M_{\mathfrak{c} m}^{s h}$ shrinks to a circle.

\vspace{.25cm}
\footnotesize 

{\fl \bf Remark:} Consistency requires that the RHS of (\ref{e:s-h-dependence-of -tht-2}) be independent of $u$, which enters through $\wp^{-1}$ and $\gamma$. We verify this by showing that $\pdr_u$(\ref{e:g-u-s-dependence-of-tht-2}) agrees with (\ref{e:u-dependence-tht-2}). In fact, from (\ref{e:g-u-s-dependence-of-tht-2}) and using $\dot{u} = \pm \sqrt{2 \la k^2 \chi(u)}$ and (\ref{e:cubic-equation-S3}),
	\beq
	\ov{\Om_2} \dd{g}{u} =  \ov{\dot{u}} - \frac{k}{f'(h)} \int_{\infty}^s s' \: \pdr_h(1/\dot{u}) \: ds' =  \ov{\dot{u}} \mp \frac{k}{f'(h)} \int_{\infty}^s \frac{\la (h + m u)}{ 2 \sqrt{2 \la k^2} (\chi(u))^{3/2}} s' \: ds'
	= \pm \frac{1+ \frac{k}{f'(h)}\left(\frac{h + m u}{2\mathfrak{c} - m^2 - \frac{2 u}{\la}}\right)}{\sqrt{2 \la k^2 \chi(u)}},
	 \eeq
which agrees with (\ref{e:u-dependence-tht-2}). As $\chi$ (\ref{e:cubic-equation-S3}) is a quadratic function of $s'$, the integrand behaves as $1/s'^2$ for large $s'$, so that the lower limit does not contribute.

\vspace{.25cm}

\normalsize

{\fl\bf Summary:} Thus, aside from the Casimirs $\mathfrak{c}$ and $m$, the action-angle variables on the union of toroidal level sets ${\cal T}$ are given by the following functions of $s$, $h$, $u$ and $\tht$:
	\beqs
	I_1 &=& \frac{k s^2}{2} + f_{\mathfrak{c} m}(h), \quad 
	\tht^1= k \left( \wp^{-1}\left(\frac{k^2 \la}{2} \left(u - \frac{\mathfrak{c} \la}{3}\right); g_2, g_3  \right) - \om_I(\mathfrak{c}, m, s, h) \right), \cr
	I_2 &=& I_2(h; \mathfrak{c}, m)  \quad \text{and} \quad
	\tht^2 = \Omega_2 \left( \frac{\tht}{f_{\mathfrak{c} m}'(h)} +  \frac{\tht^1}{k} - \ov{f_{\mathfrak{c} m}'(h)} \int^s_{\infty} s' \: \pdr_h \tht^1 \: ds' + \eta(h) \right).
	\label{e:action-angle-variables-torus}
	\eeqs
We have verified by explicit calculation that these variables are canonically conjugate. As a function of $u \in [u_{\rm min}, u_{\rm max}]$, $\tht^1$   increases from zero to $k \om_R$ (\ref{e:u-wp-function}). As noted, $\tht^2$ depends linearly on $\tht$, but finding its dependence on $u,s$ and $h$ requires the evaluation of the integral over $s'$ in (\ref{e:action-angle-variables-torus}). We have not been able to do this analytically but could evaluate it numerically for given $\mathfrak{c}$ and $m$. Here, $f, \eta$ and $I_2$ are arbitrary functions of $h$, with $f'$ and $I_2'$ non-zero and the frequency $\Om_2 = -k f'(h)/I_2'(h)$. A {\it simple choice} is to take 
	\beq
	f(h) = -I_2(h) = k h 
	\quad \text{and} \quad
	\eta(h)= 0. 
	\label{e:simple-choice-f-I2}
	\eeq
	For this choice, the Hamiltonian (\ref{e:Hamiltonian-s}) acquires a simple form in terms of the action variables
	\beq
	H = k( I_1 + I_2) + k^2 \left(\mathfrak{c} + \ov{2 \la^2} \right).
	\eeq
The corresponding frequencies $\Om_j = \pdr H/ \pdr I_j$ are then both equal to $k$. Though the frequencies are equal, the periodic coordinates $\tht^1$ and $\tht^2$ generally have different and incommensurate ranges, so that the trajectories are quasi-periodic (see Fig.~\ref{f:torus-plot-theta-phi}). While we do not have a simple formula for the range of $\tht^2$, that of $\tht^1$ is $2 k \om_R$ (twice its increment as $u$ goes from $u_{\rm min}$ to $u_{\rm max}$, see Eq. (\ref{e:u-wp-function})), which depends on the symplectic leaf and invariant torus via the four conserved quantities.

\vspace{.25cm}

{\fl \bf Relation to action-angle variables on the circular submanifold:} Finally, we show how the action-angle variables obtained above degenerate to those on the circular submanifold $\cal C$ of \S \ref{s:circular-level-sets}, where the elliptic function solutions reduce to trigonometric functions with the imaginary half-period $\om_I$ diverging. For given $\mathfrak{c}, m$ and $h$, we must let $s \to s_{\rm min}$ to reach the circular submanifold. On $\cal C$, the simple zeros of $\chi$, $u_{\rm min}$ and $u_{\rm max}$ coalesce at a double zero so that $u$ becomes a constant. Thus, the angle variable $\tht^1$ (\ref{e:action-angle-variables-torus}) ceases to be dynamical. In the same limit, from (\ref{e:action-angle-variables-torus}), the surviving angle variable $\tht^2$ becomes a linear function of $\tht$ with constant coefficients. Moreover, for the simple choices of Eq. (\ref{e:simple-choice-f-I2}), we get $I_2 = - k h$ and $\tht^2 = \tht$ upto an additive constant. Pleasantly, these action-angle variables are seen to agree with those obtained earlier on $\cal C$ (\ref{e:circular-action-variable}).


\vspace{.25cm}

{\fl \bf Acknowledgements:} We would like to thank G. Date, M. Dunajski and A. Laddha for useful discussions and references. This work was supported in part by the Infosys Foundation, J N Tata Trust and a grant (MTR/2018/000734) from the Science and Engineering Research Board, Govt. of India.

\appendix
\section{Relation to Kirchhoff's equations and Euler equations}
\label{a:Kirchhoff-equations}

Kirchhoff's equations govern the evolution of the momentum $\vec P$ and angular momentum $\vec M$ (in a body-fixed frame) of a rigid body moving in an incompressible, inviscid potential flow \cite{MT}. Here, $\vec P$ and $\vec M$ satisfy the Euclidean $\mathfrak{e}(3)$ algebra:
	\beq
	\{ M_a, M_b \} = \eps_{abc} M_c, \quad 
	\{ P_a, P_b \} = 0 \quad \text{and} \quad 
	\{ M_a, P_b \} = \eps_{abc} P_c. 
	\eeq
The Hamiltonian takes the form of a quadratic expression in $\vec P$ and $\vec M$ \cite{D-K-N}:
	\beq
	2 H = \sum a_i M_i^2 + \sum b_{ij} (P_i M_j + M_i P_j) + \sum c_{ij} P_i P_j.
	\label{e:Kirchhoff's-Hamiltonian}
	\eeq
The resulting equations of motion are 
	\beq
	\dot{\vec P} = \vec P \times \dd{H}{\vec M} \quad \text{and} \quad 
	\dot{\vec M} = \vec P \times \dd{H}{\vec P} + \vec M \times \dd{H}{\vec M}.
	\label{e:Kirchhoff's-equations}
	\eeq
Now taking $a_i =1, b_{ij} =0$ and $c_{ij} = \delta_{ij}$ and using the map $\vec M \mapsto -\vec L$ and $\vec P \mapsto \vec S - \vec K/\la$, we see that the Hamiltonian of the Kirchhoff model reduces to that of the Rajeev-Ranken model (\ref{e: H-mechanical}). However, unlike in the Kirchhoff model, $L$ and $\tl{S} = S - K/\la$ in the Rajeev-Ranken model satisfy a centrally extended $\mathfrak{e}(3)$ algebra following from Eq. (\ref{e:PB-SL-dual}):
	\beq
	\{ L_a, L_b \} = -\la \eps_{abc} L_c, \quad 
	\{ \tl{S}_a, \tl{S}_b \} = 0 \quad \text{and} \quad 
	\{ L_a, \tl{S}_b \} = -\la \eps_{abc}\left( \tl{S}_c + \frac{K_c}{\la} \right).
	\eeq
Thus, the equations of motion of the Rajeev-Ranken model (\ref{e: EOM-LS}) differ from those of the Kirchhoff model (\ref{e:Kirchhoff's-equations}). Nevertheless, this formulation implies that the equations of the Rajeev-Ranken model may be viewed as Euler-like equations for a centrally extended Euclidean algebra with the quadratic Hamiltonian $H = (L^2 + \tl{S}^2)/2$. 

Alternatively, if we use the dictionary $\vec M \mapsto -\vec L$ and $\vec P \mapsto \vec S$, then the Poisson algebras of both models are the same $\mathfrak{e}(3)$ algebra. The differences in their equations of motion may now be attributed to the linear term $\vec K \cdot \vec S/\la$ in the Rajeev-Ranken model Hamiltonian (\ref{e: H-mechanical}), which is absent in  (\ref{e:Kirchhoff's-Hamiltonian}). For more on the Kirchhoff model, its variants and their integrable cases, see for instance \cite{D-K-N, Sokolov, B-M-S}.



\begin{thebibliography}{99}

\bibitem{R-R} S. G. Rajeev and E. Ranken, {\it Highly nonlinear wave solutions in a dual to the chiral model}, Phys. Rev. D $\mathbf{93}$, 105016 (2016).

\bibitem{G-V} G. S. Krishnaswami and T. R. Vishnu, {\it On the Hamiltonian formulation and integrability of the Rajeev-Ranken model},  J. Phys. Commun. $\mathbf{3}$, 025005 (2019).

\bibitem{Z-M} V. E. Zakharov and A. V. Mikhailov, {\it Relativistically invariant two-dimensional models of field theory which are integrable by means of the inverse scattering problem method}, Zh. Eksp. Teor. Fiz. $ \mathbf{74}$, 1953 (1978).

\bibitem{Nappi} C. R. Nappi, { \it Some properties of an analog of the chiral model}, Phys. Rev. D $\mathbf{21}$, 418 (1980).

\bibitem{P-W} A. M. Polyakov and P. B. Wiegmann, {\it Theory of non-abelian Goldstone bosons in two dimensions}, Phys. Lett. B $\mathbf{131}$, 121 (1983).

\bibitem{B-T} O. Babelon and M. Talon, {\it Separation of variables for the classical and quantum Neumann model}, Nucl. Phys. B $\mathbf{379},$ 321 (1992).

\bibitem{B-B-T} O. Babelon, D. Bernard and M. Talon, {\it Introduction to classical integrable systems}, Cambridge University Press, Cambridge (2003); Chapt. 2, p. 23.

\bibitem{MT} L. M. Milne-Thomson, {\it Theoretical hydrodynamics}, $4^{\rm th}$ Ed., Macmillan, London (1962); Chapt. XVII, p. 528.

\bibitem{D-K-N} B. A. Dubrovin, I. M. Krichever and S. P. Novikov, {\it Integrable systems. I}, Encyclopaedia of mathematical sciences, 
Vol 4, Dynamical systems IV, Symplectic geometry and its applications,  V. I. Arnold and S. P. Novikov (Eds.), Springer-Verlag, Heidelberg, 173, (1990).

\bibitem{A-M} A. Yu. Alekseev, A. Z. Malkin, {\it Symplectic structure of the moduli space of flat connection on a Riemann surface}, Commun. Math. Phys. $\mathbf{169}$, 99 (1995).

\bibitem{Audin} M. Audin, {\it Lectures on Gauge theory and integrable systems}, Proceedings of the NATO advanced study institute and s\'eminaire de math\'ematiques sup\'erieures on gauge theory and  symplectic geometry, Jacques Hurtubise and Francois Lalonde (Eds.),  NATO ASI. Ser. C $\mathbf{488}$, 1 (1997).

\bibitem{F-R} V. V. Fock, A. A. Rosly, {\it Poisson structure on moduli of flat connections on Riemann surfaces and $r$-matrix}, Am. Math. Soc. Transl. $\mathbf{191}$, 67 (1999).

\bibitem{Arnold} V. I. Arnold, {\it Mathematical methods of classical mechanics}, $2^{\rm nd}$ Ed., Springer, New York (1989).

\bibitem{Milnor} J. Milnor, {\it Morse theory}, Princeton University Press, Princeton (1963).

\bibitem{G-R} I. S. Gradshteyn and I. M. Ryzhik, {\it Table of Integrals, Series and Products}, $7^{\rm th}$ Ed., Elsevier, Burlington (2007).

\bibitem{Sokolov} V. V. Sokolov, {\it A new integrable case for the Kirchhoff equation}, Theor. Math. Phys. $\mathbf{129}$, 1335 (2001).

\bibitem{B-M-S} A. V. Borisov, I. S. Mamaev, and V. V. Sokolov, {\it A new integrable case on $\mathfrak{so}(4)$},  Doklady Physics $\mathbf{46}$, 888 (2001).




\end{thebibliography}
\end{document}